\documentclass[reprint,superscriptaddress]{revtex4-2}
\usepackage{mathtools}
\usepackage{amsmath,amssymb}
\usepackage{xcolor}
\usepackage{braket}
\usepackage{graphicx}
\usepackage{epstopdf}
\usepackage{float}
\usepackage{nccmath}
\usepackage{times}
\usepackage{siunitx}
\usepackage{array}
\usepackage[normalem]{ulem} 
\usepackage{textgreek}
\usepackage{tikz-cd}
\usepackage{multirow}
\usepackage{booktabs}

\usepackage{xr-hyper} 
\usepackage{hyperref}

\usepackage[toc,page]{appendix}

\DeclareFontFamily{U}{futm}{}
\DeclareFontShape{U}{futm}{m}{n}{<-> fourier-bb}{}
\DeclareMathAlphabet{\fourierbb}{U}{futm}{m}{n}

\newcommand{\abs}[1]{\left| #1 \right|}

\newcommand{\brc}[1]{\left[ #1 \right]}

\newcommand{\ed}{\mathbf{d}}
\newcommand{\hd}{\fourierbb{H}}
\newcommand{\ded}{\fourierbb{D}}
\newcommand{\Wv}[1]{\mathbf{W}^{(#1)}}
\newcommand{\Wf}[1]{\mathcal{W}^{(#1)}}
\newcommand{\rs}{\boldsymbol{r}}

\definecolor{suggGreen}{RGB}{0,130,0}

\begin{document}

\title{A 100× Faster Beam Propagation Method for Nonlinear Optical Wave Propagation\\ Based on Discrete Exterior Calculus}






\author{Amgad Abdrabou}
\email []{amgad.abdrabou@slu.edu}
\affiliation{Department of Electrical and Computer Engineering, Saint Louis University, Saint Louis, MO 63103, USA}

\author{R. El-Ganainy}
\email[]{relganainy@slu.edu}
\affiliation{Department of Electrical and Computer Engineering, Saint Louis University,  Saint Louis, MO 63103, USA}

\begin{abstract}
Efficient simulation of nonlinear light propagation in complex photonic structures remains a major challenge because these systems combine intricate transverse geometries with propagation over distances spanning many diffraction lengths. Existing numerical methods often require computationally intensive uniform discretizations or struggle to accurately represent complex material boundaries, limiting the practical simulation of multiscale nonlinear photonic devices. Here we introduce a computational framework for solving the nonlinear Schrödinger equation that accelerates simulations by more than two orders of magnitude (over 100×) compared with conventional approaches while maintaining spectral-level accuracy, thereby reducing the computational time required to solve large-scale and multiscale problems from days to hours. The method is based on discrete exterior calculus, enabling geometry-conforming discretization directly on unstructured meshes without the weak formulations required by conventional finite-element methods. In contrast to Fourier-based spectral solvers, it avoids global oversampling, eliminates Gibbs-type oscillations at material interfaces, naturally incorporates absorbing boundary conditions, and preserves the topological structure of the underlying differential operators. A key feature of the framework is that higher-order propagation operators are generated algorithmically from lower-order discrete operators, providing a systematic route to extending simulations beyond the standard nonlinear Schr\"{o}dinger equation. Benchmarks on fundamental solitons, asymmetric beams, and optical vortices confirm spectral-level accuracy while demonstrating computational speedups exceeding 100×. By combining geometric flexibility, computational efficiency, and a systematic framework for higher-order wave propagation, our approach enables routine simulations of nonlinear dynamics in previously inaccessible multiscale photonic systems, including photonic crystal fibers, large-core multimode waveguides, and other complex integrated photonic platforms. In addition, the reported simulation speedup enables numerical optimization of nonlinear optical structures beyond what is currently feasible using simplified models or intuition alone.

\end{abstract}

\maketitle


\section{Introduction}
Nonlinear dynamics has become a central framework for understanding complex phenomena across science and engineering. Within this broad domain, nonlinear wave dynamics plays a particularly important role, governing systems in which wave evolution cannot be described by linear superposition. Such waves arise in a wide range of physical contexts: in fluid dynamics they underlie turbulence and soliton formation; in biology they describe processes such as nerve pulse propagation, cardiac dynamics, and morphogenesis; in chemical systems they appear as reaction--diffusion waves; and in atmospheric physics they govern large-scale structures such as Rossby waves and shock fronts. In optics, nonlinear wave dynamics is responsible for phenomena including self-focusing, soliton propagation, supercontinuum generation, and frequency comb formation, while in quantum systems it governs the behavior of Bose--Einstein condensates and matter-wave solitons. 

A particularly important class of nonlinear wave phenomena involves the evolution of a transverse field profile along a preferred propagation direction under a nonlinear evolution equation. Such systems exhibit rich dynamics arising from the interplay between dispersion or diffraction and nonlinearity, giving rise to phenomena including soliton formation, frequency comb generation, and supercontinuum broadening.

Among the models used to describe these systems, the nonlinear Schr\"odinger equation (NLSE) provides one of the most universal and widely applicable frameworks. It governs the evolution of slowly varying wave envelopes in nonlinear dispersive media and emerges naturally under envelope and multiple-scale approximations in the weakly nonlinear regime. In this sense, the NLSE serves as a canonical model from which a variety of other nonlinear wave equations can be systematically derived or approximated under appropriate asymptotic limits. For example, in regimes where wave dynamics is dominated by unidirectional propagation and weak nonlinearity with long-wavelength dispersion, reductions of the NLSE can lead to effective equations such as the Korteweg--de Vries (KdV) equation. More broadly, different scaling limits and perturbative expansions connect the NLSE to a hierarchy of nonlinear evolution equations, highlighting its role as a unifying framework for nonlinear wave phenomena. In nonlinear optics, the NLSE describes light propagation in nonlinear media, capturing key effects such as self-phase modulation and soliton dynamics. In quantum gases, it appears as the Gross--Pitaevskii equation, modeling the mean-field behavior of Bose--Einstein condensates. In both cases, the NLSE captures the essential interplay between longitudinal evolution and transverse structure that governs nonlinear wave propagation.

The central role of the NLSE across multiple physical systems makes its accurate and efficient numerical solution essential for modeling nonlinear wave dynamics. In this work, we focus on nonlinear optics, specifically on nonlinear wave propagation in optical fibers and waveguides, which underpins a wide range of modern technologies, including laser science, optical communications, ultrafast optics for probing chemical dynamics, and integrated photonics, just to mention a few examples. However, simulating optical wave propagation remains challenging due to the intricate interplay between nonlinearity, dispersion or diffraction, and the multiscale spatial and temporal features inherent to many practical settings. These challenges are particularly pronounced in structured optical media, where fine transverse geometric features coexist with long propagation distances, as well as in highly multimode fibers, where modal interference can generate fine spatial structures, and in strongly nonlinear regimes where effects such as beam filamentation may occur. Consequently, developing computational methods that are both efficient and capable of preserving essential physical and topological properties is of paramount importance.

To address these challenges, a variety of numerical approaches have been developed to study wave evolution under the NLSE in optical systems, including finite-difference methods (FDM), split-step methods (SSM)~\cite{weideman1986split}, and spectral methods (SM)~\cite{TahaAblowitz1984,trefethen2000spectral}, as well as more recent formulations based on the finite element method (FEM)~\cite{Zouraris2001,li2025efficient}. Among these, FDM is straightforward to implement but lacks flexibility, as the use of nonuniform meshes often degrades accuracy. Spectral and split-step methods are widely used in optics due to their efficiency for homogeneous or weakly structured systems; however, they rely on uniform grids and periodic boundary conditions, making them less suitable for complex geometries and heterogeneous media. FEM, in contrast, enables flexible meshing and can naturally accommodate geometries with multiple characteristic length scales. However, it relies on weak formulations, which become increasingly cumbersome for generalized NLSE (GNLSE) models that include multiple nonlinearities and higher-order diffraction terms. Moreover, FEM performance depends sensitively on the choice of basis functions and may suffer from stability or accuracy issues if these are not carefully selected~\cite{Chen2020Eff}. Time integration within FEM frameworks typically involves implicit schemes such as Crank--Nicolson, which can become computationally expensive in nonlinear regimes. Hybrid approaches that combine FEM for the linear step with exponential treatments of the nonlinear step alleviate some of these issues but remain computationally demanding.

In this respect, both SSM and SM have become the gold standard for simulating the GNLSE, largely thanks to the fast Fourier transform (FFT). To fully exploit the efficiency of the FFT, the transverse computational grid must be uniform, ideally comprising $2^n$ nodes in each direction or, as supported by modern FFT implementations such as FFTW, a number of nodes that factorizes into small prime numbers. While algorithms for nonuniform grids exist, they are significantly slower than uniform-grid implementations. This limitation becomes critical in multiscale problems, where different regions require different resolutions. For example, if a nonlinear wave experiences filamentation in localized regions, the grid may need to be refined only in those regions, and for an optical wave propagating in a photonic crystal fiber, which contains fine geometric structures (holes) across the transverse direction, it would be advantageous to have a fine mesh resolving these structures while keeping a coarser mesh in the core or uniform cladding. FFT-based methods are generally incapable of handling these scenarios efficiently. Another challenge is that FFT methods automatically impose periodic boundary conditions, which can produce incorrect results if part of the wave is scattered or deflected to the boundaries, as it would artificially re-enter from the opposite side.

\begin{figure}[tbph!]
    \centering
    \includegraphics[width=0.8\linewidth]{ 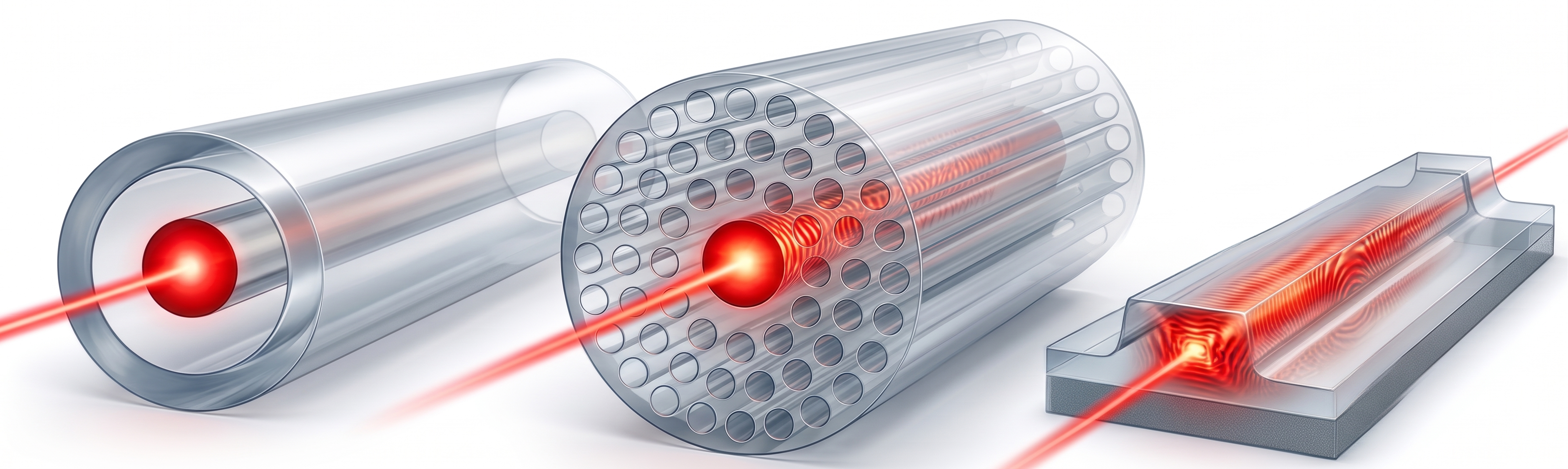}
\caption{Schematics of three representative optical waveguide platforms widely used in nonlinear optics: step-index optical fibers (left), photonic crystal fibers (center), and integrated photonic ridge waveguides (right). These examples illustrate the geometric diversity and heterogeneity of practical waveguide structures. In this work, we present a versatile computational framework based on discrete exterior calculus (DEC) that enables efficient simulation of nonlinear optical wave propagation across such geometries.}
    \label{fig:typical_waveguides}
\end{figure}

It is therefore highly beneficial to develop a computational scheme for solving the GNLSE that incorporates the following features: (1) flexible meshing to accurately handle multiscale geometries, (2) a direct discretization of the governing equation without the need to formulate a weak form, (3) seamless compatibility with absorbing boundary conditions to eliminate artifacts from scattered waves, (4) the ability to extend naturally to vectorial NLSEs, and (5) compatibility with adaptive step-size evolution schemes. Such a technique will enable fast and accurate simulation of problems involving multiple length scales, such as photonic crystal fibers or large fiber cores supporting many optical modes, both of which have recently attracted significant interest in nonlinear optics. 

In this work, we develop a computational scheme based on discrete exterior calculus (DEC) that achieves all of the aforementioned objectives. By leveraging the inherent structure-preserving properties of DEC, our approach enables flexible meshing capable of resolving multiscale geometries, direct discretization of the governing nonlinear evolution equations without the need for a weak form, and seamless integration with absorbing boundary conditions to prevent spurious reflections.  While in this work, we focus on nonlinear optical systems, the computational scheme introduced here can be applied equally well to the study of nonlinear matter waves in Bose–Einstein condensates.

\section{Results}
In this section, we present the conceptual and mathematical framework of the discrete exterior calculus beam propagation method (DEC-BPM), while deferring detailed derivations to the Supplementary Information. We then demonstrate its performance through representative examples. Before proceeding to these details, it is instructive to first provide a high-level comparison between DEC-BPM, standard spectral BPMs, and FEM-based BPMs, as summarized in Table~\ref{tab:method_comparison}. 

\begin{table*}[htbp!]
    \caption{\label{tab:method_comparison}Comparison of numerical methods for beam propagation and waveguide analysis, highlighting the structural and mathematical distinctions of the proposed DEC-BPM framework.}
    \begin{ruledtabular}
        \renewcommand{\arraystretch}{2.0}
        \begin{tabular}{l c c c}
            \textbf{Feature} & \textbf{Spectral Method} & \textbf{Finite Element Method} & \textbf{Discrete Exterior Calculus} \\
            \colrule
            \textbf{Mesh/Grid} &
            \begin{minipage}[c]{0.26\textwidth}
            \par\vspace{4pt}
            \textbf{Uniform Cartesian:} Costly for multiscale features.
                \includegraphics[width=0.9\linewidth]{ 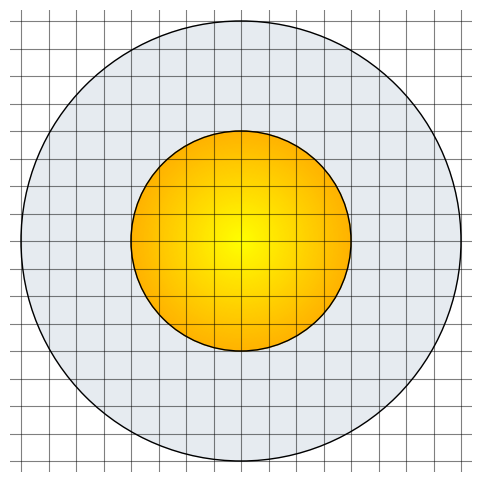}
            \end{minipage} &
            \begin{minipage}[c]{0.26\textwidth}
            \par\vspace{4pt}
                \textbf{Unstructured Primal:} Adaptable to complex geometries.
                \includegraphics[width=0.9\linewidth]{ 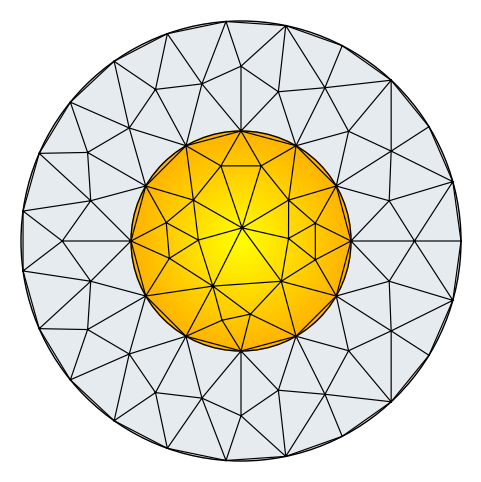}
            \end{minipage} &
            \begin{minipage}[c]{0.26\textwidth}
                \par\vspace{4pt}
                \textbf{Primal-Dual Complex}: Barycentric/Voronoi duals.
                \includegraphics[width=0.9\linewidth]{ 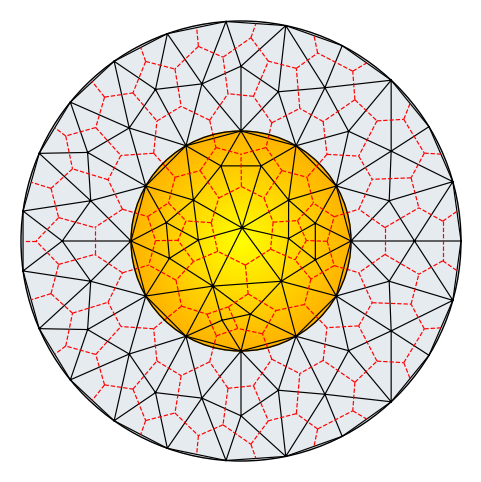}
            \end{minipage} \\
            \vspace{8pt}

            \textbf{Formulation} &
            \begin{minipage}[t]{0.26\textwidth}
                 Utilizing global basis functions, FFT-based.
            \end{minipage} &
            \begin{minipage}[t]{0.26\textwidth}
                Weak Form via shape functions.
            \end{minipage} &
            \begin{minipage}[t]{0.26\textwidth}
                Direct discretization of differential forms. Preserves exact metric-free topology ($\ed^2=0$).
            \end{minipage} \\
            \vspace{5pt}

            \textbf{Propagation} &
            \begin{minipage}[t]{0.26\textwidth}
                Typically explicit; it may use Runge-Kutta propagation, though implicit schemes are permissible.
            \end{minipage} &
            \begin{minipage}[t]{0.26\textwidth}
                Typically implicit (e.g., Crank-Nicolson via a split-step approximation). Explicit schemes require mass matrix inversion or artificial mass lumping.
            \end{minipage} &
            \begin{minipage}[t]{0.26\textwidth}
                Naturally explicit: Geometrically constructs a diagonal Hodge star $\hd_0$, enabling explicit ODE integration (e.g., via RK45) without mass matrix inversion.
            \end{minipage} \\
            \vspace{5pt}

            \textbf{Laplacian} ($\nabla^2$) &
            \begin{minipage}[t]{0.26\textwidth}
                $-k^2$ (Spectral derivative)
            \end{minipage} &
            \begin{minipage}[t]{0.26\textwidth}
                Stiffness matrix via weak form: $\int \nabla u \cdot \nabla v \, d\Omega$
            \end{minipage} &
            \begin{minipage}[t]{0.26\textwidth}
                Combinatorial incidence \& Hodge stars matrices: $-\hd_{0}^{-1} \ded_{0}^{\mathsf{T}} \hd_{1} \ded_{0}$
            \end{minipage} \\
            \vspace{5pt}

       \textbf{Higher-order} ($\nabla^{2n}$) &
            \begin{minipage}[t]{0.26\textwidth}
                $(-k^2)^n$ (Spectral derivative)
            \end{minipage} &
            \begin{minipage}[t]{0.26\textwidth}
                Requires $\boldsymbol{H}^n$-conforming elements or recursive mixed formulation splitting.
            \end{minipage} &
            \begin{minipage}[t]{0.26\textwidth}
                Cascaded matrix multiplication: $(-\hd_{0}^{-1} \ded_{0}^{\mathsf{T}} \hd_{1} \ded_{0})^n$
            \end{minipage} \\
            \vspace{5pt}

            \textbf{Accuracy} &
            \begin{minipage}[t]{0.26\textwidth}
                Exhibits Gibbs phenomenon at sharp index steps. Less optimal for complex beam profiles.
            \end{minipage} &
            \begin{minipage}[t]{0.26\textwidth}
                Accurate when explicitly aligned with material interfaces, but computationally heavier for propagation.
            \end{minipage} &
            \begin{minipage}[t]{0.26\textwidth}
                Captures sharp contrasts accurately (step-index). Highly convergent for non-standard inputs (Elliptic Gaussian, vortex beams).
            \end{minipage}
            \end{tabular}
    \end{ruledtabular}
\end{table*}

A key advantage of DEC-BPM is its natural compatibility with unstructured meshes, which enables accurate representation of complex geometries and sharp material boundaries. This flexibility also allows the computational domain to conform to the physical structure (e.g., circular or irregular cross-sections), avoiding the artificial rectangular truncation required by Cartesian grids in spectral methods. Such truncation leads to unnecessary domain padding, increased memory usage, and higher computational cost. Moreover, the DEC framework readily accommodates absorbing boundary conditions, which are essential for modeling unguided radiation and leakage in open systems. In contrast, spectral methods inherently assume periodic boundary conditions, making them ill-suited for such scenarios. 

Compared to FEM-based approaches, DEC offers a significant conceptual and practical simplification: the discretization proceeds directly from the governing equations without requiring a weak formulation. This facilitates the treatment of generalized NLSE models with multiple nonlinear terms. In addition, higher-order diffraction operators can be constructed systematically through repeated application of discrete operators associated with lower-order terms, avoiding the stringent basis function requirements typically encountered in FEM. Taken together, these features enable the use of explicit time-stepping schemes, such as fourth-order Runge--Kutta (RK4), for propagation. The compatibility of such schemes with adaptive step sizing further enhances computational efficiency, allowing faster simulations while maintaining high accuracy.

\noindent
\textbf{Conceptual Framework:---} As discussed above, we focus on modeling nonlinear optical wave propagation in guiding structures using the nonlinear Schr\"odinger equation (NLSE). In normalized form, this equation can be written as:

\begin{equation}\label{eq:NLSE_dimless}
i \frac{\partial \psi}{\partial z} = -\sum_n p_n \nabla_\perp^{2n} \psi - q_L V(x,y)\,\psi - q_{NL}\mathcal{N}(|\psi|^2)\,\psi,
\end{equation}
where $\psi(x,y,z)$ is the normalized field envelope, $z$ is the dimensionless propagation coordinate, and $\nabla_\perp^2 = \partial_x^2 + \partial_y^2$ denotes the transverse Laplacian. The terms $\nabla_\perp^{2n}$, with $n$ an integer, account for higher-order diffraction effects (typically, only the leading term is retained for weakly diffracting beams). The linear potential $V(x,y)$ represents the transverse refractive index profile of the waveguide, while $\mathcal{N}(|\psi|^2)$ describes the nonlinear response. For Kerr media, the nonlinearity takes the form $\mathcal{N}(|\psi|^2) = \chi(x,y)\,|\psi|^2$, where $\chi(x,y)$ is the Kerr coefficient, which may vary across the transverse plane. Finally, the constants $p_n$, $q_L$, and $q_{NL}$ ensure consistency of units across all terms. In practice, Eq.~\ref{eq:NLSE_dimless} is often expressed in normalized units, which we adopt after developing the formalism. Although the derivation and normalization of Eq.~\ref{eq:NLSE_dimless} are standard, they are included for completeness in Supplementary Note 1.

The first step in using DEC with any differential equation is to recast the equation in the language of exterior calculus (EC)~\cite{deschamps1981,katz1985differential,flanders1989}. A key idea in this formulation is that different physical quantities naturally live on different geometric objects. For example, a scalar field (such as refractive index or intensity) is associated with points in space, while differences of a field occur along edges, and fluxes naturally pass through areas (faces). This is not merely a mathematical construction—it reflects how these quantities are defined physically. This organization is formalized using differential forms of different orders. In an $n$-dimensional space, a $k$-form represents a quantity that is naturally associated with $k$-dimensional objects: $0$-forms correspond to pointwise values (scalars), $1$-forms to quantities along edges, $2$-forms to quantities over surfaces, and so on up to $n$-forms, which are associated with volumes. The order of the form therefore reflects both the type of physical quantity and the dimensionality of the domain it interacts with. This viewpoint provides a unified way of describing different physical quantities based on how they interact with the geometry of the domain. The above discussion illustrates, at a conceptual level, why DEC in two dimensions employs two complementary meshes (see table \ref{tab:method_comparison}. Each mesh represents different types of quantities that naturally reside on distinct geometric elements, such as edges or faces.

\begin{figure}[t]
    \centering
\includegraphics[width=0.9\columnwidth]{ 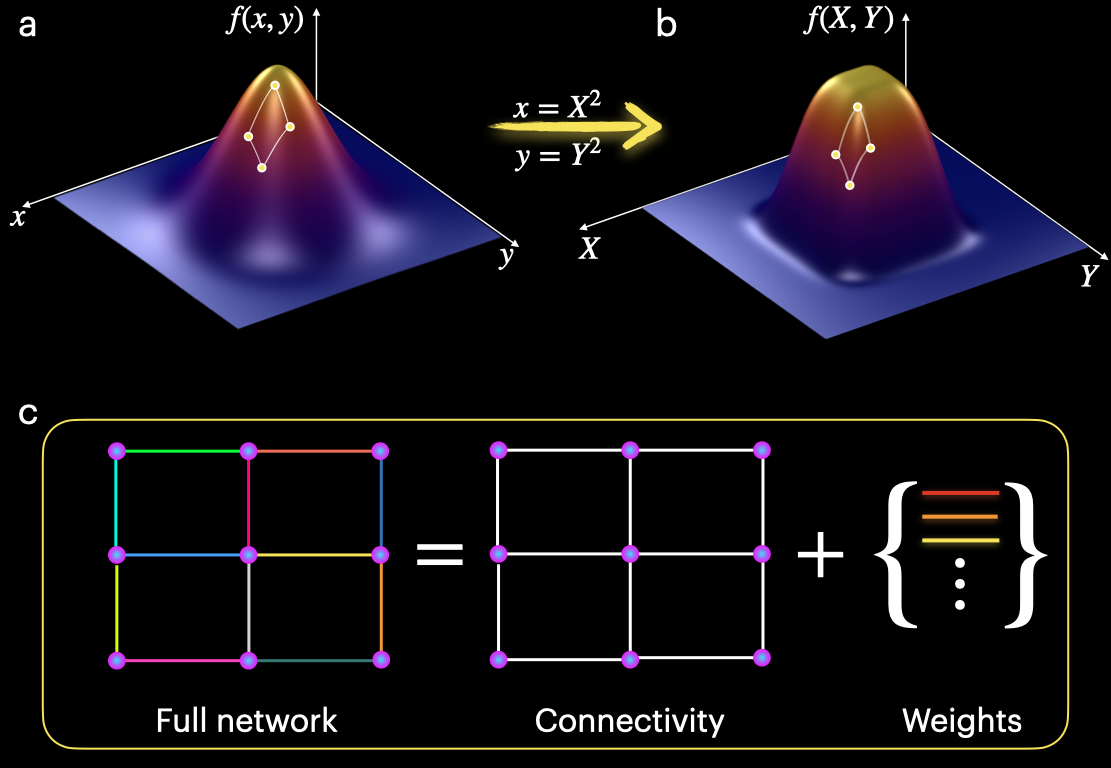}
    \caption{(a) A scalar function $f(x,y)$ and (b) its mapped representation under the coordinate deformation $x=X^2$ and $y=Y^2$. Although the scalar field itself remains unchanged, its gradient (or directional derivative along any given direction) depends on the local coordinate system. Nevertheless, the integral quantity $\oint df$ around any closed loop, such as those shown in the figures, always vanishes. This reflects a topological property of the connected vertices (white dots) that is independent of the local coordinate deformation. (c) Illustration of the same concept in discrete networks. The vertices are connected through weighted edges, with the weights encoded by the edge colors. Equivalently, the network can be represented by an unweighted connectivity graph (white edges), which encodes the topology, together with a separate list of edge weights. One of the main advantages of DEC over many other numerical schemes is that it naturally separates topology from geometry. As a result, identities and conservation laws of the original continuous problem, such as $\mathrm{curl}\,\mathrm{grad}\,f = 0$, remain exactly preserved after discretization.  }
    \label{fig:Topo}
\end{figure}

Next, the various operations—whether differential operators or linear and nonlinear mappings—must be expressed in terms of a set of standard exterior calculus (EC) operators that relate differential forms of different orders. These operators are defined in a way that separates the topology of the problem from its metric properties. One of these fundamental operators is the exterior derivative, $\ed$, which unifies the notions of gradient (when acting on $0$-forms), curl (when acting on $1$-forms), and divergence (when acting on $2$-forms) into a single, coordinate-independent operation. Importantly, this operator is purely topological, meaning it depends only on how points are connected, rather than on distances or angles. Figure \ref{fig:Topo} illustrates this concept from both continuous and discrete perspectives. Panel (a) shows the scalar function $f(x,y)$ in the original coordinates $(x,y)$, while panel (b) depicts the same function in the deformed coordinate system $(X,Y)$. Although the scalar field itself is unchanged, its gradient and directional derivatives generally differ in the two coordinate systems. Nevertheless, the integral quantity $\oint df$ around any closed loop always vanishes. In numerical analysis, one is often interested in approximating quantities such as $\nabla f$. However, many discretization schemes do not guarantee that the identity $\nabla \times \nabla f = 0$ is exactly preserved, even though it should hold independently of the chosen coordinate system or deformation. In contrast, discrete exterior calculus (DEC), through the definition of the exterior derivative operator $\ed$, preserves this topological structure by construction, regardless of the coordinate representation. Panel (c) further illustrates this idea using a discrete network representation, where vertices are connected by weighted edges encoded by their colors. Conceptually, the information contained in the network can be separated into two parts: the connectivity of the graph (topology) and the edge weights (metric information). Within the framework of DEC, the operator $\ed$ isolates precisely this topological information. To recover physically meaningful quantities, such as gradients per unit length or fluxes per unit area, the geometric information of the domain must be incorporated. This is achieved through the Hodge star operator, $\star$, which is metric-dependent and accounts for lengths, areas, and volumes in the system. The Hodge star operator plays a complementary role by connecting these different types of quantities through geometric information such as lengths, areas, and volumes. In particular, it maps a $k$-form into an $(n-k)$-form, allowing conversion between quantities defined on complementary geometric objects (for example, between edge-based and face-based quantities). 

In DEC these ideas are implemented directly on a computational mesh. The exterior derivative, $\ed$, is represented exactly using matrices that describe how mesh elements are connected—for example, which edges connect to which nodes, or which faces are bounded by which edges. These matrices act as discrete versions of derivatives, capturing how quantities change across the mesh. On the other hand, the Hodge star operator, which depends on geometry, is approximated using matrices that encode the size and shape of the mesh elements. These matrices enable consistent conversion between quantities defined on different parts of the mesh, ensuring that both the geometry and the underlying physical structure of the problem are properly captured. It is important to emphasize that, in the context of DEC, the separation between topological and metric operators serves a purpose far beyond mere mathematical elegance. As discussed earlier, this separation guarantees that key structural properties of the underlying continuous problem---such as the identity $\nabla \times \nabla f = 0$---are preserved exactly at the discrete level. More generally, DEC ensures that the fundamental relation $d^2 f = 0$, which encompasses identities of this kind, is always satisfied on the discrete mesh of the computational domain. This stands in contrast to standard discretization schemes, such as finite differences (FD), finite elements (FEM), or spectral methods, where such identities are typically satisfied only approximately.\\

\noindent
\textbf{Mathematical Formulation of the EC-NLSE:---} Having outlined the conceptual framework of EC, we now present the basic mathematical formulation for representing the NLSE in the EC language. However, since EC and DEC, along with their associated concepts such as forms and differential forms, are not widely familiar in the context of nonlinear optics, and to preserve the flow of the manuscript, we restrict ourselves here to introducing only the essential concepts, notations, and final formulations. A more detailed mathematical treatment is deferred to  Supplementary Note 2. Before proceeding, we make three important remarks. First, we emphasize that EC is a framework rather than a rigid representation of the underlying PDE (the NLSE in our case). Consequently, there is more than one way to define the variables and to cast the PDE in the language of exterior calculus. Second, once a particular choice is made, it is essential to ensure that all terms in the resulting formulation correspond to the same $k$-form. In this sense, the requirement is analogous to enforcing consistency of physical units across all terms. Third, while the operators $\ed$ and $\star$ are defined abstractly for differential forms of any degree, their explicit action depends on both the degree $k$ of the form and the dimensionality of the space, $n$. For example, applying $\ed$ to a $0$-form (a scalar function defined at points) produces a $1$-form (quantities associated with edges), corresponding to the gradient operator. When applied to a $1$-form, $\ed$ yields a $2$-form (quantities associated with oriented areas), corresponding to the curl operator. In general, the exterior derivative maps a $k$-form to a $(k+1)$-form. On the other hand, the Hodge star operator $\star$ maps a $k$-form to an $(n-k)$-form. \\

In this work, we begin by noting that the optical field is described by a scalar function $\psi(x,y,z)$, which we naturally treat as a $0$-form. Similarly, the scalar potential is also represented as a $0$-form. Other scalar quantities, such as $\chi(x,y)$ and $|\psi|^2$, are likewise treated as $0$-forms. Within the DEC framework, these quantities are therefore defined on the vertices of the primal mesh. Their multiplication is understood as pointwise multiplication, which, in the language of EC, corresponds to the wedge product $\wedge$ between $0$-forms. Next, as shown in Supplementary Note 2.1, the Laplacian operator can be expressed as $\star \ed \star \ed$. Taken together, these considerations lead to the following exterior calculus formulation of the NLSE:

\begin{equation}\label{eq:NLSE_EC1}
i \frac{\partial \psi}{\partial z} = -  \star \ed \star \ed \psi -  V\wedge \,\psi - \, \chi \wedge |\psi|^2 \wedge \,\psi.
\end{equation}
In the above equation, we retain only the leading-order diffraction term and consider Kerr nonlinearity, both of which are relevant to many practical applications. The more general form of the equation, including higher-order terms, is presented in Supplementary Note 2.2. For clarity, we also set $p_1=q_L=q_{NL}=1$, although other normalizations can be adopted straightforwardly.

Note that every term in this equation is consistently a $0$-form, as expected. While this formulation is mathematically equivalent to the original equation, it is not ideally suited for numerical implementation. The primary reason is that the potential function may be discontinuous for sharply varying profiles, making its value less well-defined at boundaries. The exterior calculus formalism provides a systematic framework for addressing this problem by representing
the linear potential and nonlinear response using pulse basis functions defined on the triangles,
rather than in terms of delta functions at the nodes~\cite{bossavit1998computational,Bossavit1999}. In other words, the potentials are defined as a constant value per triangle. In particular, by applying the $\star$ operator to Eq.~\ref{eq:NLSE_EC1} from the left, and using the property that applying $\star$ twice yields the identity, we obtain:

\begin{equation}\label{eq:NLSE_EC2}
i \star \frac{\partial \psi}{\partial z} = - \ed \star \ed \psi - \star V\wedge \,\psi -  \star \, \chi \wedge |\psi|^2 \wedge \,\psi.
\end{equation}

In this form, every term in the equation is a $2$-form, which can be interpreted as representing an averaged quantity over the area associated with each vertex (see Supplementary Note 2.3 for details). This interpretation naturally introduces a form of local averaging, improving the numerical robustness of the formulation, particularly in the presence of sharp spatial variations. Next, since the quantities \(V\), \(\psi\), \(\chi\), and \(|\psi|^2\) are all \(0\)-forms, the above equation can be recast as (see Supplementary Note 2.4 for more details): 
\begin{equation}\label{eq:NLSE_EC2}
i \star \frac{\partial \psi}{\partial z}
= -  \ed \star \ed \psi
-  \star_V \psi
-  \star_{\mathcal{N}} \psi ,
\end{equation}
where \(\star_V\) and \(\star_\mathcal{N}\) are linear operators associated with the potential \(V\) and the nonlinear interaction \(\chi |\psi|^2\), respectively. The role of these operators will become clearer when we discuss the discrete formulation of this equation below.\\

\begin{figure*}[t]
\centering
\includegraphics[width=0.8\linewidth]{ 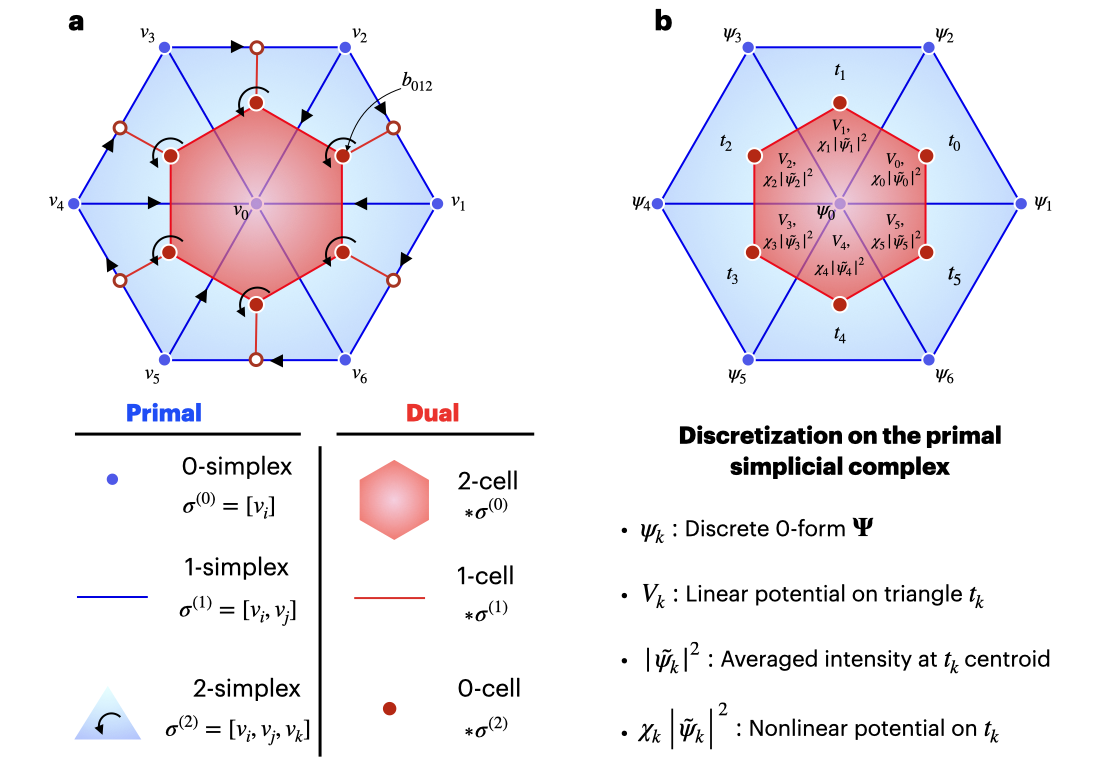}
\caption{(a) Geometric representation of a two-dimensional primal simplicial complex and its barycentric dual. The top panel shows a local mesh patch surrounding a central primal vertex, $v_0$. The shaded red region denotes the corresponding dual $2$-cell, constructed by connecting the barycenters of the neighboring primal $2$-simplices (e.g., $b_{012}$) through the barycenters (midpoints) of the shared primal $1$-simplices. The lower panel summarizes the duality mapping induced by the discrete Hodge star operator, $\ast$, which maps each primal $k$-simplex, $\sigma^{(k)}$, to a dual $(2-k)$-cell, $\ast \sigma^{(k)}$. In this two-dimensional complex, the $2$-simplices are assigned a consistent counterclockwise orientation determined by the ordered vertices $[v_i, v_j, v_k]$, whereas the orientation of the $1$-simplices (edges) is chosen arbitrarily. (b) Discretization of the potential and nonlinear terms on the primal simplicial complex. 
Each triangle $t_k = [v_{k_1}, v_{k_2}, v_{k_3}]$ carries a piecewise-constant refractive-index potential $V_k$ and a 
nonlinear contribution $\chi_k\,\abs{\tilde{\psi}_k}^2$, where 
$\abs{\tilde{\psi}_k}^2= \bigl(|\psi_{{k_1}}|^2 + |\psi_{{k_2}}|^2 + |\psi_{{k_3}}|^2\bigr)/3$ 
is the averaged intensity over the three vertices $ [v_{k_1}, v_{k_2}, v_{k_3}]$ of $t_k$.}
\label{fig:Mesh}
\end{figure*}

\noindent
\textbf{DEC computational framework:---}
Next, we present the DEC computational framework employed in this work. As discussed earlier, different physical quantities are represented on different elements of the DEC computational mesh. To achieve this, DEC utilizes complementary primal and dual meshes~\cite{hirani2003discrete,desbrun2005discrete,grady2010,Abdrabou2026hybrid}, as illustrated schematically in Fig.~\ref{fig:Mesh}(a) for a two-dimensional computational domain. 

The blue mesh represents the primal mesh, whose vertices, edges, and areas correspond to 0-, 1-, and 2-simplices, respectively. These simplices host the primal (i.e., independent and not derived from one another) 0-, 1-, and 2-forms in the same order. The red mesh, on the other hand, represents the dual mesh, which similarly consists of its own 0-, 1-, and 2-cells. The connection between the primal and dual meshes is established through the Hodge star operator. For example, applying the Hodge star operator to a primal 0-form (i.e., a quantity defined on the blue vertices) produces a dual 2-form that resides on the corresponding dual 2-cell, represented by the red-shaded hexagonal region. This primal–dual pairing separates topology, encoded combinatorially by the exterior derivative on the primal mesh, from geometry, encoded metrically by the Hodge star through ratios of primal and dual cell volumes or via the Galerkin process.

More rigorously, the simplicial complex, denoted by $\mathcal{K}$, contains a natural hierarchy of simplices: 2-simplex faces (triangles), their 1-simplex edges, and their 0-simplex nodes. Each $k$-simplex $\sigma^k \in \mathcal{K}$, for $k = 0, \ldots, 2$, is defined by its $k+1$ vertices as $\sigma^k = [v_1, \ldots, v_{k+1}]$, where the subscripts denote node indices \cite{hirani2003discrete}. The 0-simplices are the vertices $\{v_1, \ldots, v_{N_0}\}$, where $N_0$ is the number of nodes. The 1-simplices are edges $\{e_1, \ldots, e_{N_1}\}$, with each edge $e_k \equiv \sigma^1_k = [v_{k_1}, v_{k_2}]$ oriented according to the ordering of its endpoints, and $N_1$ is the number of edges. The 2-simplices are triangles $\{t_1, \ldots, t_{N_2}\}$, with each triangle $t_k \equiv \sigma^2_k = [v_{k_1}, v_{k_2}, v_{k_3}]$ defined by three nodes and bounded by three 1-simplex edges. The orientation of the 2-simplices is assumed to be consistent throughout the mesh (e.g., counterclockwise in Fig.~\ref{fig:Mesh}) (a), a condition typically enforced by mesh generation tools. For lower-dimensional simplices ($k < 1$), we adopt a canonical orientation: each edge $[v_{k_1}, v_{k_2}]$ is oriented such that $k_1 < k_2$. Additionally, the orientation of each primal k-simplex induces a compatible orientation on the corresponding dual $(n-k)$-cell~\cite{desbrun2005discrete}.\\

\noindent
\textbf{\textit{Discrete exterior derivative}:-} Since the simulation domain is represented by a simplicial complex $\mathcal{K}$, the smooth differential forms must also be discretized. In DEC, this is achieved by integrating each smooth differential $k$-form over the corresponding $k$-simplices of the mesh. The resulting discrete spaces are denoted by $\mathcal{C}^k(\mathcal{K})$, which represent discrete $k$-forms defined on $\mathcal{K}$.

The discretization is carried out through the de Rham map $\mathcal{R}$. For example, the primal $0$-form $\psi$ and the $1$-form $\ed\psi$ are discretized as:

\begin{equation}\label{eq:zerofrom}
\mathcal{R}(\psi)=\boldsymbol{\Psi} =
\left[\psi(v_1),\ldots,\psi(v_{N_0})\right]^\mathsf{T}
\in \mathcal{C}^0(\mathcal{K}),
\end{equation}

and

\begin{equation}\label{eq:onefrom}
\mathcal{R}(\ed\psi)=
\left[
\int_{e_1}\ed\psi,\ldots,
\int_{e_{N_1}}\ed\psi
\right]^\mathsf{T}
\in \mathcal{C}^1(\mathcal{K}),
\end{equation}

where $\mathsf{T}$ denotes the transpose. While Eq.~\eqref{eq:zerofrom} simply evaluates $\psi$ at the mesh vertices, Eq.~\eqref{eq:onefrom} requires the discrete exterior derivative, denoted by $\ded_k$, which maps discrete $k$-forms to discrete $(k+1)$-forms.

We note that the operator $\ded_k$ is purely combinatorial and depends only on the topology of the mesh. Moreover, it is represented by a sparse $(N_{k+1}\times N_k)$ incidence matrix encoding the signed relations between $k$-simplices and $(k+1)$-simplices. Its construction follows directly from the generalized Stokes' theorem,
\begin{equation}\label{eq:stokes}
\int_{\partial \mathcal{M}} \boldsymbol{\alpha}
=
\int_{\mathcal{M}} \ed \boldsymbol{\alpha},
\end{equation}
where $\boldsymbol{\alpha}$ is a $k$-form and $\mathcal{M}$ is an oriented $(k+1)$-dimensional manifold with boundary $\partial\mathcal{M}$.

For the case of the smooth $0$-form $\psi$, the exterior derivative $\ed\psi$ is a $1$-form integrated along each oriented edge $e_k=[v_{k_1},v_{k_2}]$. Applying Stokes' theorem gives
\[
[\mathcal{R}(\ed\psi)]_k
=
\int_{e_k}\ed\psi
=
\psi(v_{k_2})-\psi(v_{k_1})
=
\sum_\ell [\ded_0]_{k\ell}\psi_\ell,
\]
where $\psi_\ell=\psi(v_\ell)$ and $k=1,\ldots,N_1$. The matrix $\ded_0\in\mathbb{R}^{N_1\times N_0}$ therefore maps vertex-based quantities to edge-based quantities, with entries
\[
[\ded_0]_{k\ell}=
\begin{cases}
+1, & \text{if node $\ell$ is the head of edge $k$},\\
-1, & \text{if node $\ell$ is the tail of edge $k$},\\
0, & \text{otherwise}.
\end{cases}
\]

Higher-order discrete exterior derivatives, such as $\ded_1$ (and $\ded_2$ in three dimensions), are constructed analogously.\\

\noindent
\textbf{\textit{Discrete Hodge star}:-} Certain operations---notably the Hodge star, which maps $k$-forms to $(n-k)$-forms, where $n$ is the dimension of $\mathcal{K}$---are often conceptually associated with a dual complex $\ast\mathcal{K}$. In the specific case of Delaunay meshes~\cite{DelaunayHodge}, the dual complex is constructed using Voronoi cells~\cite{Voronoi2007}, ensuring that primal and dual elements are mutually orthogonal. While the discrete exterior derivative $\ded_k$ is purely topological, the discrete Hodge star $\hd_k$ encodes the metric structure of the domain. In our 2D problem, the Hodge star operator maps primal $k$-forms to dual $(2-k)$-forms by relating the measure of a primal $k$-simplex to that of its dual $(2-k)$-cell.  

For general unstructured meshes, the primal-dual mesh orthogonality is not preserved. To address this, we adopt the Galerkin formulation for the Hodge star using Whitney forms~\cite{whitney1957}. The mapping is defined via the variational formulation~\cite{Bossavit1999}, which ensures numerical stability and consistency on general meshes without requiring the explicit geometric construction of an orthogonal dual mesh.  

The Galerkin Hodge star operator is denoted here by $\hd_k$. In general, we only need the Galerkin approach for $\hd_1$, leading to a sparse non-diagonal matrix. The operators $\hd_0$ and $\hd_2$ can be obtained using the barycentric dual mesh, and they are \emph{diagonal} matrices with positive entries~\cite{mohamed2016}. This enables efficient explicit time integration or propagation, since the inverse of $\hd_0$ is trivial. To better explain how this procedure is applied to our problem, let us recall Eq.~\eqref{eq:NLSE_EC2}:
\[
i \star \frac{\partial \psi}{\partial z} = - \mathbf{d} \star \mathbf{d} \psi -  \star_V \psi - \star_{\mathcal{N}} \psi.
\]
To discretize the above equation, four distinct discrete Hodge star operators need to be defined. They correspond to the $\star$ operator on the left-hand side of the equation, $\star$ operator on the right-hand side (which are different because they act on 0-forms and 1-forms, respectively), as well as the operators $\star_V$ and $\star_{\mathcal{N}}$. Their discrete representations are denoted by $\hd_0$, $\hd_1$, $\hd_0(V)$, and $\hd_0(\mathcal{N})$, respectively. The operators $\hd_0$ and $\hd_1$ are purely geometrical, i.e., only constructed via the mesh metrics (triangle areas or edge lengths). On the other hand, operators $\hd_0(V)$ and $\hd_0(\mathcal{N})$ encode the information about the linear and nonlinear potential, respectively. For the linear potential $V$, we assign to each triangle $t_k$ a value $V_k$ of the potential evaluated at the centroid of triangle $t_k$. The nonlinear response $\mathcal{N}$ is also defined as $\chi_k\,\abs{\tilde{\psi}_k}^2$, where 
$\abs{\tilde{\psi}_k}^2:= \bigl(|\psi_{{k_1}}|^2 + |\psi_{{k_2}}|^2 + |\psi_{{k_3}}|^2\bigr)/3$ 
is the averaged intensity over the three vertices $ [v_{k_1}, v_{k_2}, v_{k_3}]$ of triangle $t_k$. These details are also presented in Fig~\ref{fig:Mesh}(b) and its caption. The entries of the diagonal matrices $\hd_0$ are then given by~\cite{mohamed2016}
\begin{equation*}
\brc{\hd_0}_{p,p} =  \frac{1}{3}\sum_l\abs{t_{p_l}},\quad 1\leq p \leq N_0, 
\end{equation*}
where $\set{t_{p_l}}_{l=1}^{n_p}$ is the set of triangles $n_p$ connected to the node $p$ and $\abs{t_{p_l}}$ is the area of the triangle $t_{p_l}$.
The factor $1/3$ accounts for the portion of the triangle area that is added to the total effective dual area for node $p$. Similarly, the entries of the diagonal matrices $\hd_0(V)$ and $\hd_0(\chi\abs{\psi}^2)$ are given by
\begin{equation*}
 \brc{\hd_0(V)}_{p,p} =  \frac{1}{3}\sum_l V_{p_l}\abs{t_{p_l}},\quad 1\leq p \leq N_0,
\end{equation*}
and
\begin{equation*}
 \brc{\hd_0(\chi \abs{\psi}^2)}_{p,p} =  \frac{1}{3}\sum_l  \chi_{p_l}\abs{\tilde{\psi}_{p_l}}^2\abs{t_{p_l}},\quad 1\leq p \leq N_0,
\end{equation*}
Finally, the Hodge operator $\hd_1$ is constructed via Whitney 1-forms. Given a generic triangle $t$ with nodes denoted locally as $[v_1,v_2,v_3]$, a Whitney 0-form $\Wf{0}_\ell$ is associated with each node $v_\ell,\, \ell = 1,\ldots, 3$, and is given by
\begin{equation*}
\Wf{0}_{v_\ell} = \lambda_\ell(\rs),
\end{equation*}
where $\lambda(\rs)$ is the barycentric coordinates for the triangle $t$. Since 0-forms are scalars, there is no distinction between the Whitney 0-forms and their proxy fields~\cite{lohi2021}. The Whitney 1-forms are associated with edges. The locally indexed edges of the triangle $t$ are $e_\ell, \,\ell = 1,\ldots,3$, where the edge $e_\ell$ is defined by the nodes forming it as $e_l = [v_{l_1},v_{l_2}]$. The Whitney 1-form associated with the edge $e_\ell$ is
\begin{equation*}
\Wf{1}_{e_\ell} = \lambda_{l_1}(\rs)\ed \lambda_{l_2}(\rs)-  \lambda_{l_2}(\rs)\ed \lambda_{l_1}(\rs),
\end{equation*}
while its proxy field $\Wv{1}_{e_\ell}$ is
\begin{equation}\label{eq:Whitney1Proxy}
    \Wv{1}_{e_\ell} = \lambda_{l_1}(\rs)\nabla \lambda_{l_2}(\rs)-  \lambda_{l_2}(\rs)\nabla \lambda_{l_1}(\rs). 
\end{equation}
The entries of the Galerkin Hodge star $\hd_1$ are then given by~\cite{Bossavit1999}
\begin{equation}\label{eq:ghs1}
  \brc{\hd_1}_{p,q} = \int_{\Omega} \xi\,\Wv{1}_{e_p}\cdot\Wv{1}_{e_q} d\Omega,\quad 1 \leq p,q \leq N_1
\end{equation}
\noindent
\textbf{DEC beam propagation:---}
Having outlined how to construct the key discrete operators needed for our work, we now write the DEC form of the NLES: 
\begin{equation}\label{eq:ODE_form}
i \frac{\partial \mathbf{\Psi}}{\partial z} =\boldsymbol{\mathcal{L}} \mathbf{\Psi} + \boldsymbol{\mathcal{N}}(\abs{\psi}^2) \mathbf{\Psi},
\end{equation}
where the operators $\boldsymbol{L}$ and $\boldsymbol{\mathcal{N}}(\abs{\psi}^2)$ are given, respectively, by
\begin{equation}\label{eq:linop}
\boldsymbol{\mathcal{L}} = -\hd_0^{-1} \left[ \ded_0^\mathsf{T} \hd_1 \ded_0 + \hd_0(V)\right],
\end{equation}
and
\begin{equation}\label{eq:nonlinop}
\boldsymbol{\mathcal{N}}(\abs{\psi}^2) = -\hd_0^{-1} \hd_0 (\chi |\mathbf{\psi}|^2).
\end{equation}
A key computational advantage of the DEC formulation is that $\hd_0$ is diagonal, making the computation of $\hd_0^{-1}$ particularly inexpensive. Consequently, the resulting numerical scheme naturally lends itself to \emph{explicit} time-integration and beam-propagation methods. An important implementation detail is that the nonlinear term requires only \emph{one} sparse matrix-vector product per stage. First, the values of the $\abs{\tilde{\psi_i}}^2$ at triangles $i =1..N_2$ are found at each step using a pre-computed projection operator $\mathbf{Q}$ that maps vertex-based values to element-based values by averaging.
\begin{equation}\label{eq:Q_matrix}
    \mathbf{Q} = \frac{1}{6} |\ded_1| |\ded_0|.
\end{equation}
Second, notice that the transpose $\mathbf{Q}^\mathsf{T}$ is an $(N_0 \times N_2)$ matrix that maps element-based values back to vertex-based values by accumulating contributions from all adjacent triangles, which is indeed the operation we need to accumulate dual area portions and element-wise potential values of the triangles that share a given node. So, in matrix notation, the discrete nonlinear operator in Eq.~\eqref{eq:nonlinop} is given by
\begin{equation}\label{eq:nonlinear_element}
    \boldsymbol{\mathcal{N}}(\abs{\psi}^2) = -\hd_0^{-1} \text{diag}\big(\mathbf{Q}^\mathsf{T} \mathbf{D} \mathbf{Q} |\mathbf{\Psi}|^2\big) \mathbf{\Psi},
\end{equation}
where $\mathbf{D} = \text{diag}(\chi_i |t_i|)$ is an $(N_2 \times N_2)$ diagonal matrix with entries $(\chi_i |t_i|)$ for each triangle $t_i$. Notice that the combined action of calculating the nonlinear potential and the construction of the Hodge star operator $\hd_0(\chi\abs{\psi}^2)$ is efficiently computed via the pre-computed sparse matrix $\mathbf{Q}^\mathsf{T} \mathbf{D} \mathbf{Q}$, given in Eq.~\eqref{eq:nonlinear_element}. 

For beam propagation of both methods, we employ the adaptive-step Runge–Kutta–Fehlberg (RK45) scheme~\cite{dormand1980family}, as implemented in \texttt{scipy.integrate.solve\_ivp}. The RK45 method is a fifth-order accurate explicit integrator with an embedded fourth-order error estimator, allowing for automatic step-size control based on a user-specified tolerance. In all of our simulations, and for both the DEC and spectral methods,  we chose a fixed relative tolerance of \texttt{rtol} = $10^{-9}$ and an absolute tolerance of \texttt{atol} = $10^{-11}$ for the RK45 solver. For well-conditioned triangular meshes, both $\boldsymbol{\mathcal{L}}$ and $\boldsymbol{\mathcal{N}}(\abs{\psi}^2)$ are sparse matrices with total non-zero entries $n_{nz}$, yielding a computational complexity of $\mathcal{O}(n_{nz})$ per stage. This linear scaling, combined with the adaptive step-size control, enables highly efficient propagation over long distances without sacrificing accuracy. This is significantly more efficient than split-step and spectral methods, which typically require multiple FFT operations per step (each scaling as $\mathcal{O}(n \log n)$ for $n$ grid points), or implicit FEM schemes, which necessitate solving a nonlinear system at each step via iterative methods such as Newton–Raphson, incurring substantial additional overhead.\\

\noindent
\textbf{Numerical Benchmarking and Performance:---}
We now evaluate the performance of the proposed DEC-BPM framework in terms of accuracy, convergence, and computational efficiency by benchmarking it against the gold-standard FFT-based algorithm. In particular, we compare it with the Spectral Method in the Interaction Picture (SIP), as detailed in Supplementary Note 3, which is widely regarded as one of the most computationally efficient methods for solving the nonlinear Schrödinger equation. 

The geometry considered throughout this work is a standard step-index fiber. This choice is motivated both by its role as the canonical waveguide geometry in nonlinear optics applications and by the stringent numerical challenge posed by its discontinuous refractive-index profile, which is difficult to represent accurately using many conventional methods. The physical and computational parameters used in the simulations are summarized in Table~\ref{tab:sim_params}. For computational efficiency, all simulations are performed using the standard normalized form of the NLSE (see Supplementary Note 1 for details).

To provide a comprehensive assessment across a broad range of conditions, we consider several benchmark scenarios of progressively increasing geometric and topological complexity:

\begin{enumerate}
    \item \textbf{Fundamental soliton:-} A stationary nonlinear eigenmode is propagated over long distances to evaluate physical fidelity and the preservation of conserved quantities.
    
    \item \textbf{Off-axis elliptic Gaussian beam:-} A non-eigenmode excitation with broken cylindrical symmetry is used to examine geometric robustness in the presence of a sharp core--cladding interface.
    
    \item \textbf{Elliptic optical vortex:-} A beam carrying orbital angular momentum ($m = 1$), centered on the fiber axis, is considered to assess the method's ability to resolve higher-order modal distributions that interact strongly with the core--cladding interface.
\end{enumerate}

To ensure a fair comparison, all simulations employ the same adaptive RK45 integrator (from \texttt{SciPy}) to integrate the equation of motion along the propagation direction. Under this common integration protocol, the adaptive step size is governed by the spectral radius of each transverse discretization (the largest eigenvalue magnitude of the matrix representing the linear differential operator) rather than by the propagation physics alone (i.e. geometry and initial condition). The reported wall-clock times therefore reflect both the accuracy achieved per degree of freedom \emph{and} the stiffness that each spatial discretization presents to an explicit integrator, both of which are intrinsic properties of the underlying numerical scheme.

\begin{table}[htbp!]
    \centering
    \caption{Physical and computational parameters used in all test cases.}
    \label{tab:sim_params}
    \renewcommand{\arraystretch}{1.2}
    \begin{tabular}{lcc}
        \hline
        \textbf{Parameter} & \textbf{Symbol} & \textbf{Value} \\
        \hline
        \multicolumn{3}{c}{\textit{Fiber Properties}} \\
        Operating wavelength        & $\lambda$              & $1.55\,\mu\text{m}$ \\
        Core radius                 & $R_0$                  & $25.0\,\mu\text{m}$ \\
        Normalization constant along $z$ & $L_d$          & $7347.27\mu\text{m}$\\
        Core linear index           & $n_{\text{core}}$      & $1.450$ \\
        Cladding linear index       & $n_{\text{clad}}$      & $1.449$ \\
        Core nonlinear index        & $n_{2,\text{core}}$    & $2.8\times10^{-20}\,\text{m}^2/\text{W}$ \\
        Cladding nonlinear index    & $n_{2,\text{clad}}$    & $2.2\times10^{-20}\,\text{m}^2/\text{W}$ \\
        \hline
        \multicolumn{3}{c}{\textit{Computational Domain (normalized to $R_0$)}} \\
        SIP & \multicolumn{2}{c}{$[-4,4]\times[-4,4]$} \\
        DEC & \multicolumn{2}{c}{$\{(x,y)\in\mathbb{R}^2 \mid \sqrt{x^2+y^2}\le 4\}$} \\
         \hline
      
    \end{tabular}
\end{table}

Crucially, to validate and benchmark the proposed method, reliable reference solutions must first be established. For stable soliton propagation, the analytical soliton solution provides a natural benchmark. In scenarios where analytical solutions are unavailable, reference solutions are instead generated numerically using highly resolved simulations with the SIP method, the gold-standard FFT-based solver for the nonlinear Schrödinger equation. Specifically, the reference solutions are constructed as follows:
 
\begin{itemize}
    \item \textbf{Soliton Case:} We compute the fundamental soliton eigenmode on a dense ($1024 \times 1024$) grid using a self-consistent nonlinear eigenmode solver, as described in the Supplementary Note 4. Because this is a stationary self-consistent solution with an invariant intensity profile, the reference intensity is simply the input beam profile.

    \item \textbf{Elliptic Gaussian and Vortex Cases:} For these two examples, the nonlinear Schrödinger equation is propagated using the SIP method on a dense ($512 \times 512)$ grid, and the resulting field distribution at the target propagation distance is taken as the reference solution. To keep the computational cost of generating these reference solutions manageable, and because of the SIP method's inability to accurately handle radiated fields over longer propagation distances, we restrict the benchmark to a maximum normalized propagation distance of ($z=0.1$).
\end{itemize}

We will refer to the reference solutions discussed above as $\psi_{\mathrm{ref}}(\mathbf{r},z)$ 
 and we will use the relative $L_2$-norm as a measure of the error:
\begin{equation}
E_{\mathrm{rel}}(z)
=
\frac{
\lVert |\psi(\mathbf{r},z)|^2-|\psi_{\mathrm{ref}}(\mathbf{r},z)|^2\rVert_2
}{
\lVert |\psi_{\mathrm{ref}}(\mathbf{r},z)|^2\rVert_2
},
\label{eq:error_metric}
\end{equation}
Importantly, we note that all reported wall-clock times, including mesh generation and initialization overheads, were measured on a 32-core AMD Ryzen Threadripper Pro workstation equipped with 256 GB of RAM.

\noindent
\textbf{\textit{Fundamental soliton}:-} As mentioned earlier, our first example considers the propagation of a soliton in a standard optical fiber geometry (see Table~\ref{tab:sim_params}). Since the soliton is a stationary solution of the NLSE, any distortion of its shape during propagation provides a direct measure of numerical error. This example therefore serves as a stringent test of the DEC method.

\begin{figure}[htbp!]
    \centering
    \includegraphics[width=\linewidth]{ 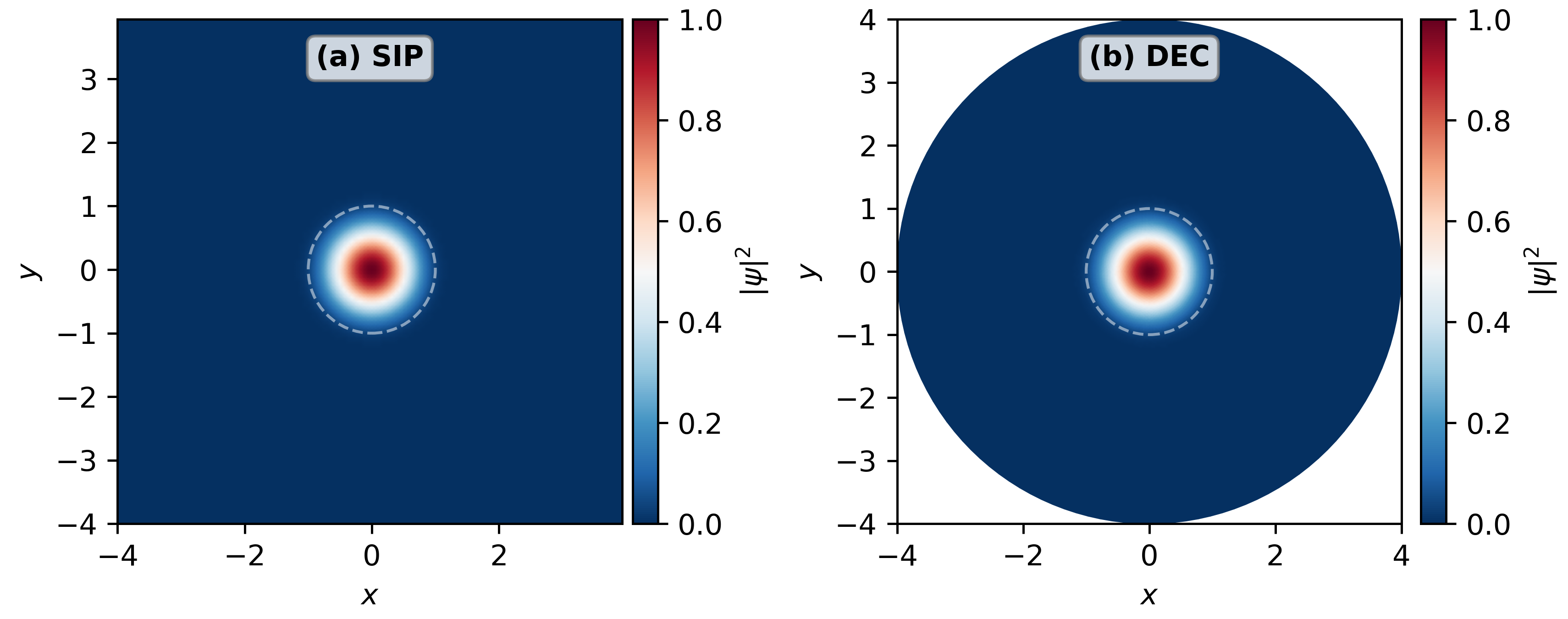}
    \caption{Initial intensity profile of soliton solutions obtained via the Spectral and DEC methods.}
    \label{fig:InputSolitonBeams}
\end{figure}

\begin{figure}[htbp!]
    \centering
    \includegraphics[width=\linewidth]{ 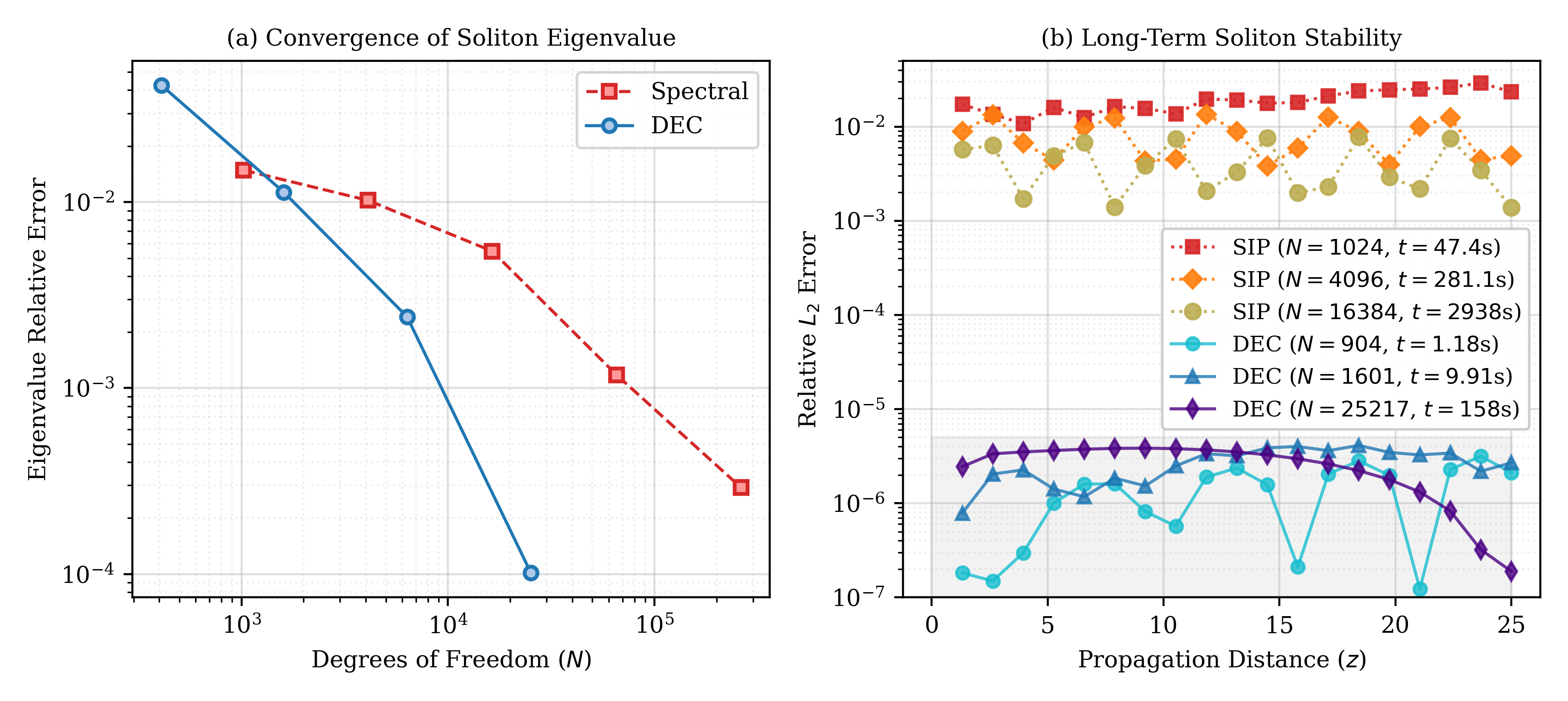}
    \caption{Fundamental soliton propagation in a step-index fiber.
    \textbf{(a)} Convergence of the soliton eigenvalue $\mu$ versus degrees
    of freedom $N$. The DEC method (blue circles) reaches agreement
    with the reference value using $\approx 10\times$ fewer unknowns than the
    SIP method (red squares), whose convergence is slowed by the staircase
    representation of the circular core--cladding interface on the uniform
    Cartesian grid.
   	\textbf{(b)} Long-term shape-preservation error $E_{\text{rel}}$ over a
		propagation distance $z = 25$. SIP fluctuates in the $10^{-3}$--$10^{-2}$, whereas DEC
		maintains fidelity near the numerical noise floor ($\sim 10^{-6}$).}
    \label{fig:SolitonBenchmark}
\end{figure}

The input soliton profiles used in the SIP and DEC simulations are shown in Fig.~\ref{fig:InputSolitonBeams}. These profiles were obtained numerically using the self-consistent method described in Supplementary Note 4. Each soliton is computed self-consistently on the same grid or mesh subsequently used for its propagation, so that the launch field is a discrete eigenmode of the operator that evolves it. Figure~S1 in the Supplementary Note 5 compares the transverse propagation of the soliton simulated via the SIP and DEC methods across varying spatial resolutions. Figure~\ref{fig:SolitonBenchmark}(a) presents the convergence of the computed soliton eigenvalue $\mu$ as a function of the number of degrees of freedom $N$ for both methods. 

 As a reference, we compute the soliton eigenvalue $\mu_{\mathrm{ref}}=-3.5537$ using the SIP solver on a $1024\times1024$ reference grid. The two solvers are mutually consistent at the $10^{-3}$ level, with the DEC solver reaching this agreement using approximately an order of magnitude fewer unknowns ($\mu_{\mathrm{DEC}}=-3.5540$ at $N=25{,}217$ versus $\mu_{\mathrm{SIP}}=-3.5547$ at $N=262{,}144$). This discrepancy in performance is primarily attributed to the sharp material discontinuity introduced by the step-index fiber profile, which induces Gibbs oscillations in Fourier-based spectral methods and consequently slows their convergence. By contrast, the second-order DEC formulation naturally accommodates local discontinuities without introducing global oscillatory artifacts, enabling it to accurately capture the step-index profile while achieving substantially higher computational efficiency.

To study the propagation dynamics, we consider a normalized propagation distance of $z=25$, corresponding to a physical propagation length of $25L_d = 18.37$ cm. The long-term stability advantage of the DEC approach is illustrated in Fig.~\ref{fig:SolitonBenchmark}(b), which shows the evolution of the relative error $E_{\text{rel}}$ as given by Eq.~\eqref{eq:error_metric} over the entire propagation distance. Throughout the simulation, the error associated with the SIP approach oscillates between $10^{-3}$ and $10^{-2}$. In contrast, the DEC method preserves the soliton profile with errors remaining near the numerical noise floor ($\sim 10^{-6}$) over the entire propagation range, demonstrating superior long-term stability and shape-preserving fidelity.

\begin{figure}[htbp!]
    \centering
    \includegraphics[width=\linewidth]{ 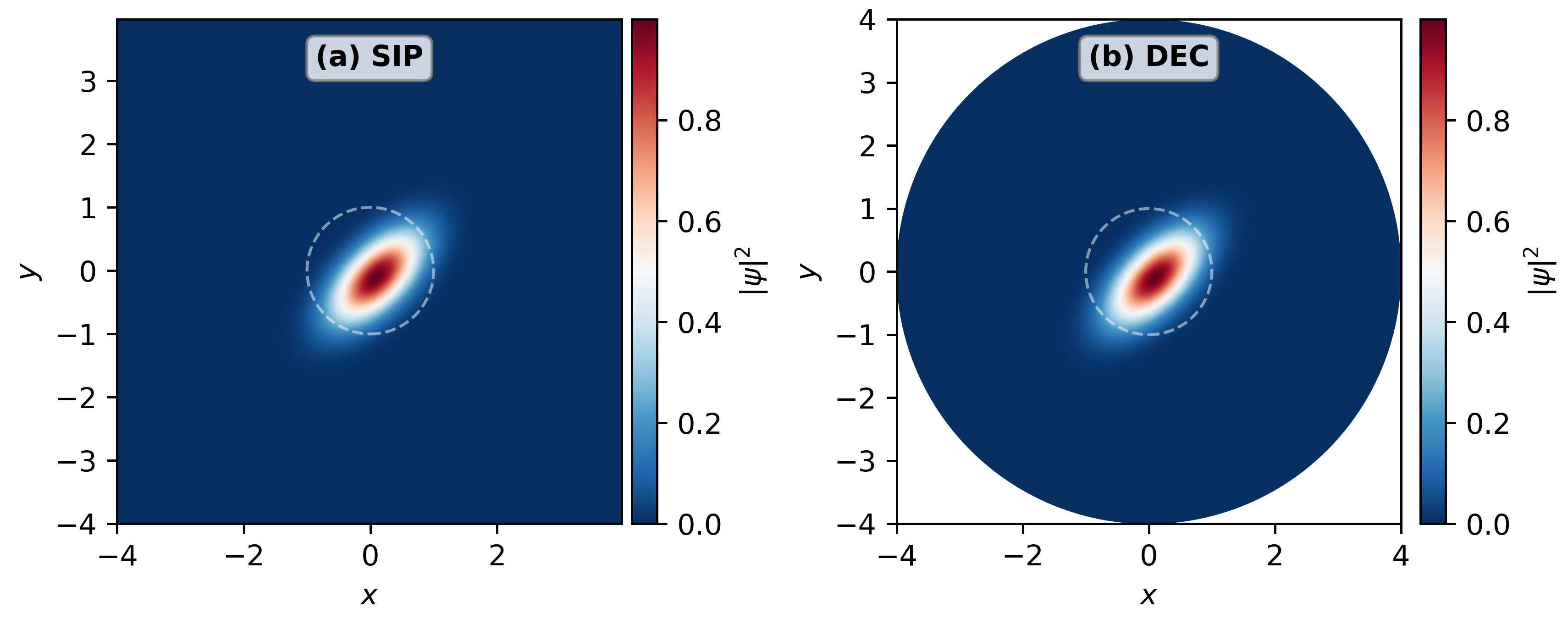}
    \caption{Initial transverse intensity profiles ($|\psi|^2$) of the off-axis elliptic Gaussian beam in normalized units. \textbf{(a)} Initialization on the Cartesian grid for the SIP method. \textbf{(b)} Initialization on the unstructured circular domain for the DEC method. The beam is shifted off-center to $(x_0, y_0) = (0.1, -0.1)$, rotated by an angle of $\theta = \pi/4$, and exhibits an elliptical aspect ratio of $w_x/w_y = 2$. The dashed white circle designates the core--cladding interface ($r = R_0$) of the step-index fiber.}
    \label{fig:InputBeamGaussian}
\end{figure}

\noindent
\textbf{\textit{Off-axis elliptic Gaussian beam}:-} The second test case is specifically designed to stress-test the ability of each numerical method to accurately handle sharp material discontinuities in the presence of a dynamically evolving and spatially asymmetric optical field. To this end, we launch an off-axis elliptic Gaussian beam. Since this input profile is not an eigenmode of the step-index waveguide, it undergoes pronounced beam breathing and self-phase modulation during propagation, causing the optical energy to repeatedly interact with the core--cladding interface. The initial field envelope is given by:
\begin{equation}
   \psi_0(U,V) = A\exp\!\left(-\frac{U^2}{w_x^2} - \frac{V^2}{w_y^2}\right),
\end{equation}
where $(U,V)$ represent the rotated transverse coordinates centered at the beam's offset position $(x_0, y_0)$ and inclined by an azimuthal angle $\theta$. These are defined through the spatial transformation:
\begin{equation}
    \begin{aligned}
        U &=(x - x_0)\cos\theta + (y - y_0)\sin\theta, \\
        V &=-(x - x_0)\sin\theta + (y - y_0)\cos\theta.
    \end{aligned}
    \label{eq:UV_coords}
\end{equation}
In our simulations, we used the following parameters: amplitude $A = 1.0$, asymmetric beam waists $w_x = 1.2$ and $w_y = 0.6$, tilt angle $\theta = \pi/4$, and an off-axis shift of $(x_0,y_0)=(0.1,-0.1)$. The resulting transverse intensity profiles, $|\psi(x,y,z=0)|^2$, initialized on both the Cartesian spectral grid and the unstructured DEC mesh, are shown in Fig.~\ref{fig:InputBeamGaussian}. By intentionally breaking the cylindrical symmetry of the waveguide, this configuration provides a stringent test of the unstructured-mesh capabilities of the DEC solver.

\begin{figure}[htbp!]
    \centering  \includegraphics[width=\linewidth]{ 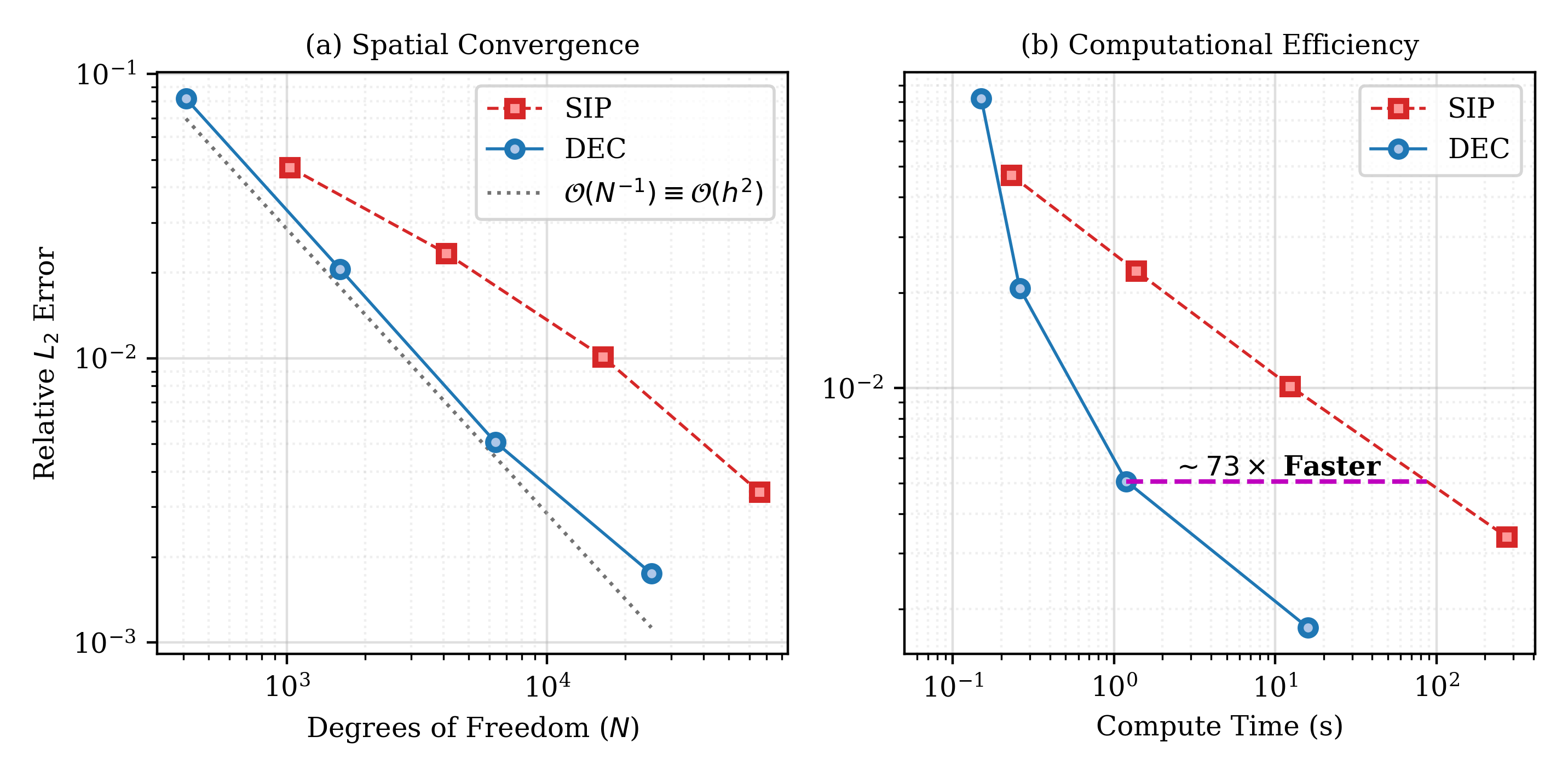}
    \caption{Performance benchmark for the off-axis elliptic Gaussian beam in a step-index fiber.
    \textbf{(a)} Relative $L_2$ error versus degrees of freedom $N$. The DEC
    method (blue circles) maintains second-order convergence $\mathcal{O}(h^2)$,
    corresponding to the observed $\mathcal{O}(N^{-1})$ scaling. Conversely, the
    SIP method (red squares) is limited to first-order $\mathcal{O}(h^1)$
    convergence (observed here as $\mathcal{O}(N^{-0.5})$) by the Gibbs phenomenon 
    at the refractive-index discontinuity.
    \textbf{(b)} Error versus compute time. At the representative accuracy
    threshold $E_{\text{rel}} = 5\times10^{-3}$ (dashed magenta line), DEC requires
    $\approx 1.19$ s versus $\approx 87.15$ s for
    SIP, a speedup of 73x.}
    \label{fig:EllipticalGaussian}
\end{figure}
\begin{figure*}[htbp!]
    \centering
    \includegraphics[width=\linewidth]{ 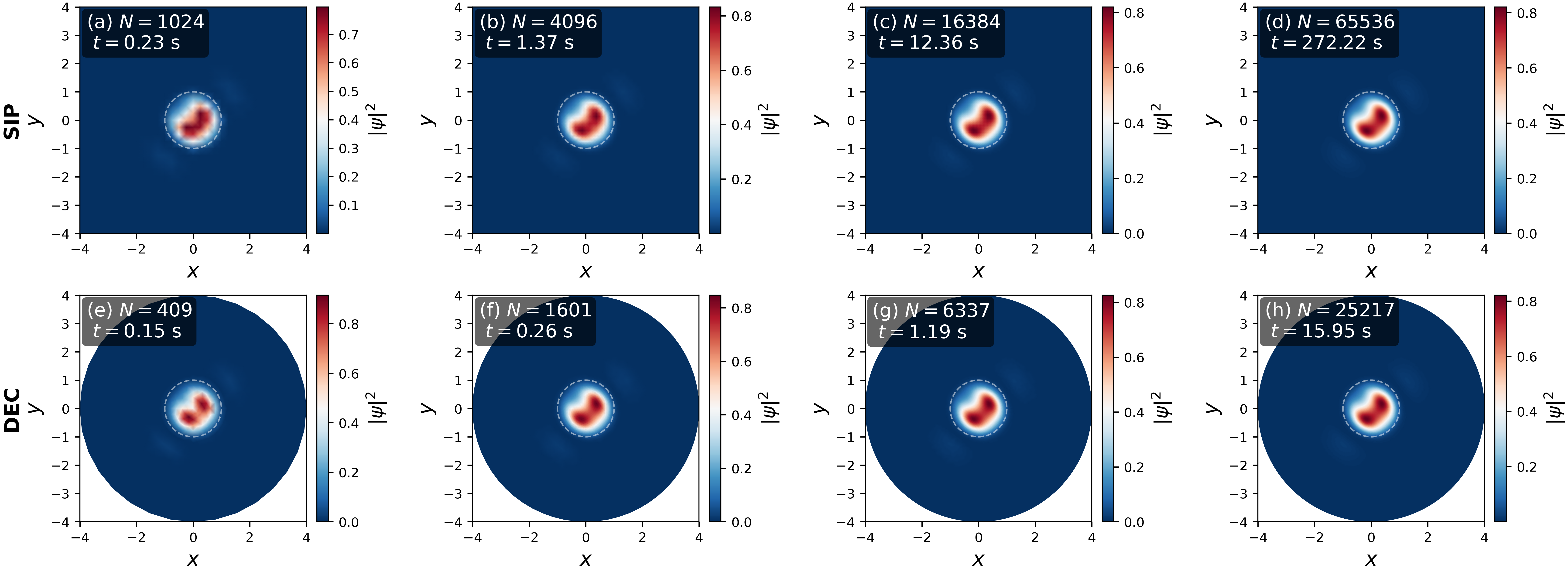}
    \caption{Visual comparison of the computed transverse intensity profiles 
    ($|\psi|^2$) for the off-axis elliptic Gaussian beam at the dimensionless 
    propagation distance of $z = 0.1$. \textbf{(Top row, a--d)} Solutions 
    obtained using the SIP method on a square Cartesian grid, 
    illustrating the grid density required to control Gibbs ringing at the refractive-index discontinuity. \textbf{(Bottom row, e--h)} Corresponding solutions 
    using the DEC method on an unstructured circular domain. Each subplot is 
    annotated with the total number of degrees of freedom ($N$) and the required 
    compute time ($t$) to reach $z = 0.1$. The dashed white circle denotes the 
    core-cladding interface of the step-index fiber. The DEC method achieves a comparable high-resolution field representation to the spectral approach but 
    with substantially fewer unknowns, highlighting its superior computational 
    efficiency for discontinuous refractive-index profiles.}
    \label{fig:GaussianGridMesh}
\end{figure*}

To compare the performance of the SIP and DEC methods, we plot the convergence behavior and computational cost of both approaches for this input beam in Fig.~\ref{fig:EllipticalGaussian}(a)--(b), respectively. The DEC formulation consistently outperforms the SIP solver, requiring a factor of 6.2 fewer degrees of freedom to achieve comparable accuracy and yielding a computational speedup of approximately 73 times. More specifically, the DEC method exhibits clear second-order convergence, $\mathcal{O}(h^2)$, corresponding to an $\mathcal{O}(N^{-1})$ scaling with respect to the number of degrees of freedom $N$. This behavior is confirmed numerically as the DEC mesh is refined from $N=409$ to $25{,}217$ nodes, for which the measured convergence order steadily improves from 2.04 to 2.15, while the relative error decreases to $1.09 \times 10^{-3}$. In contrast, the SIP method suffers from Gibbs oscillations induced by the sharp refractive-index discontinuity at the core--cladding interface, which degrades its theoretical exponential convergence to an effective first-order behavior, $\mathcal{O}(h^1)$$~$ (observed numerically as $\mathcal{O}(N^{-0.5})$), consistent with the $L_2$ representation error of a discontinuous coefficient sampled on a uniform grid. As a result, the spectral solver remains severely bottlenecked across the tested resolutions, causing its computational cost to increase rapidly with accuracy demands. For instance, achieving a representative relative error threshold of $5 \times 10^{-3}$ requires approximately 87.15 seconds using the SIP solver, whereas the DEC method---which is naturally insensitive to local discontinuities---reaches the same accuracy in only 1.19 seconds. This corresponds to a $73$-fold speedup under the common-integrator protocol, highlighting both the efficiency and robustness of the DEC framework for modeling optical fibers with step-index profiles.

The qualitative behavior underlying the convergence results is illustrated in Fig.~\ref{fig:GaussianGridMesh}, which shows the transverse intensity profiles, $|\psi|^2$, of the input beam after propagation to the normalized distance $z=0.1$ for different numbers of degrees of freedom (grid points for the SIP method and mesh elements for the DEC method). At this stage of propagation, the beam has already begun to interact strongly with the step-index interface while undergoing nonlinear reshaping. The top row corresponds to the SIP solution on a Cartesian grid, whereas the lower row shows the DEC solution on an unstructured circular mesh conforming to the fiber geometry. Figure~S2 in Supplementary Note 5 compares the transverse propagation dynamics of the elliptic Gaussian beam simulated via the SIP and DEC methods across increasing spatial resolutions.

Although both methods capture the overall confinement of the optical field within the fiber core (marked by the dashed circle), the SIP solution exhibits pronounced Gibbs ringing near the sharp refractive-index discontinuity unless extremely fine grids are employed. Consequently, achieving acceptable accuracy with the SIP solver requires a substantial increase in computational resolution and runtime, even for relatively short propagation distances. In contrast, the DEC discretization naturally conforms to the material interface and therefore produces smooth, artifact-free field distributions using only a small fraction of the degrees of freedom. This geometric adaptability enables the DEC framework to achieve high-fidelity solutions with significantly reduced computational cost.

\begin{figure}[htbp!]
    \centering
    \includegraphics[width=\linewidth]{ 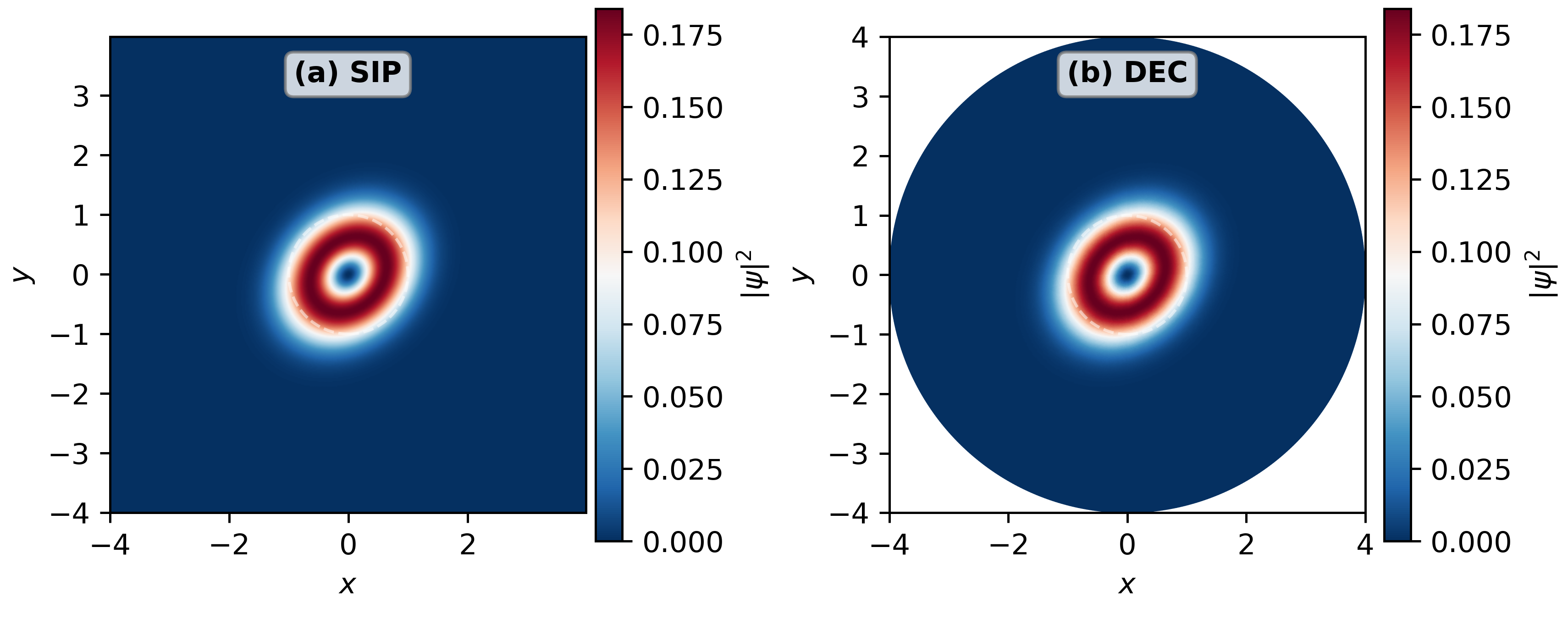}
    \caption{Initial transverse intensity profiles ($|\psi|^2$) of the elliptic optical vortex beam 
    in normalized units. \textbf{(a)} Initialization on the Cartesian grid for the SIP method. 
    \textbf{(b)} Initialization on the unstructured circular domain for the DEC method. 
    The beam is centered at the origin $(0,0)$, rotated by $\theta = \pi/4$, and carries 
    an orbital angular momentum (OAM) with topological charge $m = 1$.}
    \label{fig:InputBeamVortex}
\end{figure}

\textbf{\textit{Elliptic optical vortex}:-} Next, we consider the third example: an elliptic optical vortex beam carrying orbital angular momentum (OAM) with topological charge $m=1$. To construct the elliptic vortex beam, we again employ the rotated coordinate system defined in Eq.~\eqref{eq:UV_coords}, now centered on the fiber axis, $(x_0,y_0)=(0,0)$, with rotation angle $\theta=\pi/4$. The input field profile is given by:
\begin{equation}
    \psi_0(U,V)
    =
    A \rho(U,V) e^{i\Phi(U,V)}
    \exp\!\left(
        -\frac{U^2}{w_x^2}
        -\frac{V^2}{w_y^2}
    \right),
\end{equation}
where the complex prefactor has been expressed in polar form with an elliptical amplitude $\rho(U,V) = \sqrt{(U/w_x)^2 + (V/w_y)^2}$ and a spatially varying phase $\Phi(U,V) = \arg\left(\frac{U}{w_x} + i\frac{V}{w_y}\right)$. This representation explicitly highlights the characteristic phase winding associated with the vortex structure. Here, we set $A = 1.0$, $w_x=1.0$ and $w_y=0.8$. Figures~\ref{fig:InputBeamVortex}(a) and (b) illustrate the transverse intensity profile of this beam on both the Cartesian spectral grid and the unstructured DEC mesh prior to propagation.
 
\begin{figure}[htbp!]
    \centering
    \includegraphics[width=\linewidth]{ 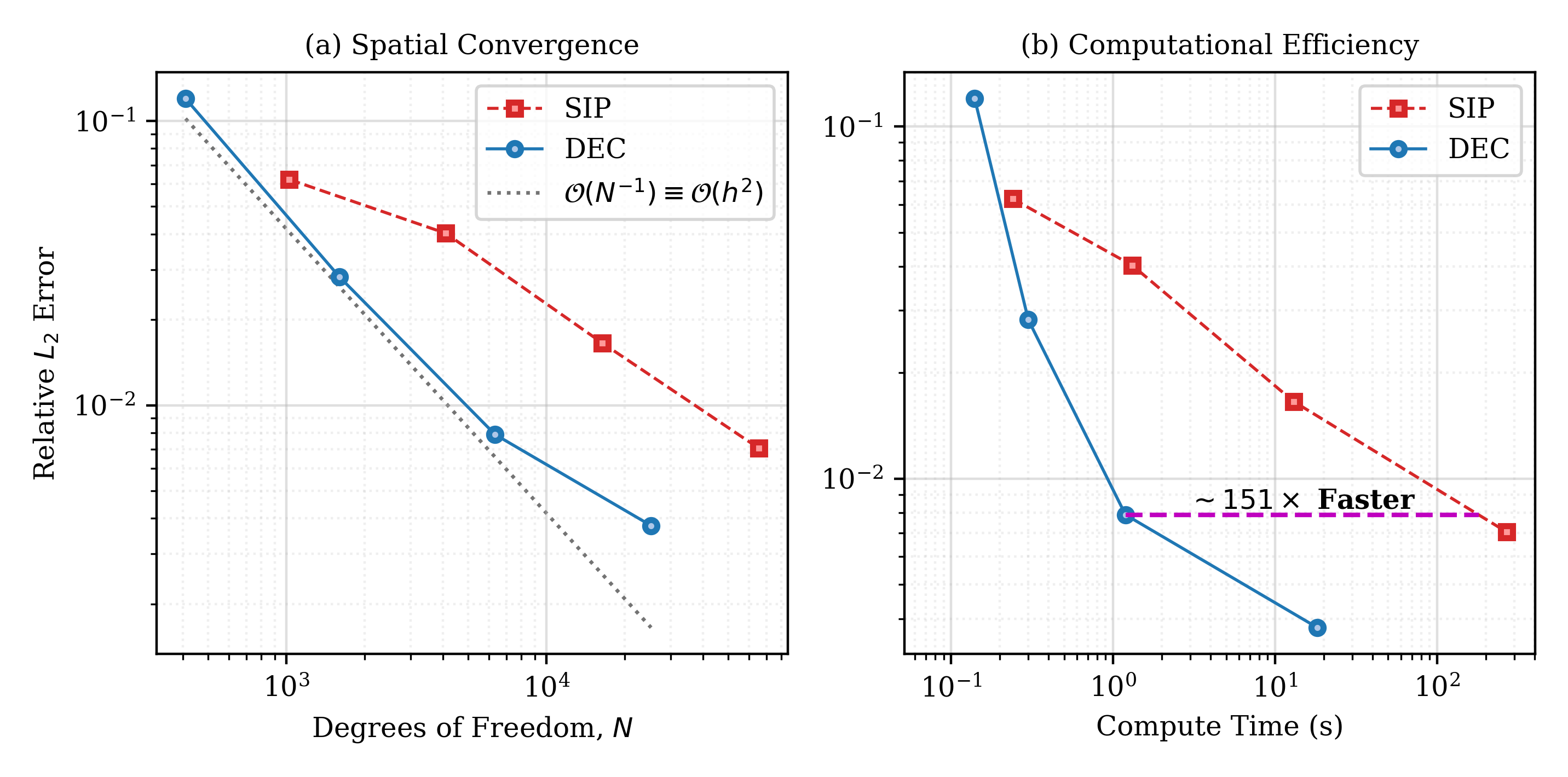}
    \caption{Performance benchmark for the elliptic optical vortex ($m = 1$). 
    \textbf{(a)} Relative $L_2$ error versus degrees of freedom $N$. DEC 
    (blue circles) maintains second-order convergence $\mathcal{O}(h^2)$ that corresponds to the observed $\mathcal{O}(N^{-1})$ scaling, while the SIP error (red squares) decreases non-monotonically, consistent with the resolution-dependent staircase bias of the discrete interface.
    \textbf{(b)} Error versus compute time. At the accuracy threshold 
    $E_{\text{rel}} = 8\times10^{-3}$ (dashed line), DEC requires 
    $\approx 1.2$ s; interpolation of the measured SIP curve gives a speedup of $\approx 150\times$ under the common RK45 protocol.}
    \label{fig:Vortex_Convergence}
\end{figure}

Having defined the input profile, we next propagate the beam in both solvers over a normalized distance of $z=0.1$. Figure~\ref{fig:Vortex_Convergence}(a) and (b) compare the convergence behavior and computational efficiency of the two numerical schemes. From the plots, it is evident that the DEC method maintains a highly stable second-order convergence rate, $\mathcal{O}(h^2)$, corresponding to an $\mathcal{O}(N^{-1})$ scaling with respect to the number of degrees of freedom. In particular, as the DEC mesh is refined from $N=409$ to $25{,}217$ nodes, the measured convergence order remains consistently between 2.02 and 2.17, while the relative error decreases to $1.56\times10^{-3}$ in only 18.28 seconds. Conversely, the SIP method initially exhibits a noticeably shallow convergence trajectory at lower resolutions. Although its convergence rate improves as the spectral grid is further refined, the method remains significantly less efficient overall. At its highest resolution of $N = 65{,}536$, the spectral solver takes nearly 269 seconds only to reach a relative error of $7.06\times10^{-3}$. Figure~\ref{fig:Vortex_Convergence}(b) further highlights this disparity at the accuracy threshold $E_{\text{rel}} = 8\times10^{-3}$: the DEC method reaches this target in $\approx 1.2$~seconds, whereas interpolation along the SIP curve---whose finest point, $7.06\times10^{-3}$ in $269$~seconds, already exceeds this accuracy---gives $\approx 181$~seconds at the threshold, a speedup of $\approx 150\times$ under the common-integrator protocol.

The widening of the performance gap relative to the Gaussian case calls for care in attribution. The vortex \emph{field} itself is smooth: $\psi_0 \propto (U/w_x + iV/w_y)\exp(-U^2/w_x^2 - V^2/w_y^2)$ is an entire function of the transverse coordinates, and the on-axis phase singularity coincides with an intensity null at which the field is locally linear---neither discretization has intrinsic difficulty representing it. The enhanced gap instead originates at the material interface. First, the $m=1$ launch projects predominantly onto odd-azimuthal guided modes, including the near-cutoff LP$_{31}$ group (cutoff $V = 5.14$ against the fiber's $V = 5.46$), whose fields concentrate at and beyond the core--cladding boundary; the geometric error of the staircased interface is therefore sampled precisely where the field resides. Second, the vortex launch has poorer overlap with the guided-mode set and sheds a larger radiated fraction than the Gaussian; on the uniform spectral grid, this halo must be represented at full resolution over the entire rectangular domain, whereas the graded DEC mesh resolves the core finely and the outer region coarsely. Third, the $\mathcal{O}(h)$ variation of the staircased core radius produces the non-monotone SIP convergence noted above.

\begin{figure*}[hbpt!]
    \centering
    \includegraphics[width=\linewidth]{ 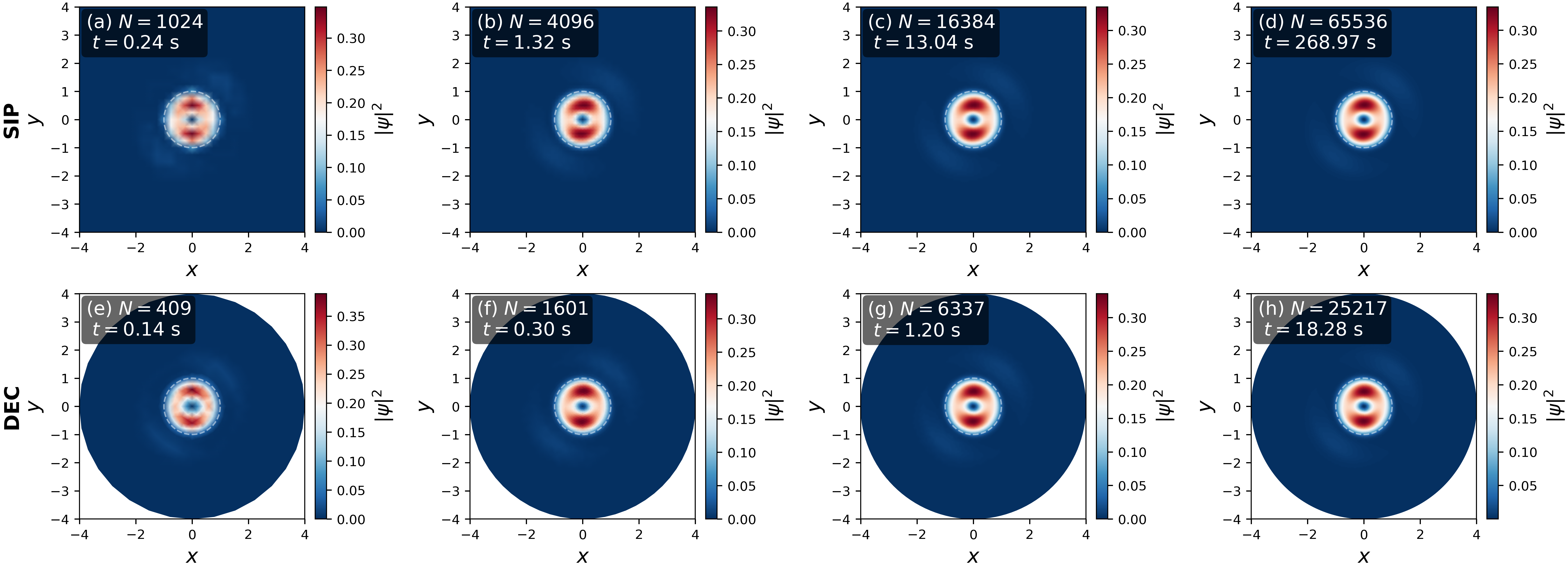}
    \caption{Visual comparison of the computed transverse intensity profiles 
    ($|\psi|^2$) for the elliptic optical vortex ($m = 1$) at the dimensionless 
    propagation distance of $z = 0.1$. \textbf{(Top row, a--d)} Solutions 
    obtained using the SIP method on a square Cartesian grid, illustrating the grid density required to control Gibbs errors at the step-index interface. \textbf{(Bottom row, e--h)} Corresponding 
    solutions using the DEC method on an unstructured circular domain. Each 
    subplot is annotated with the total number of degrees of freedom ($N$) and the 
    required compute time ($t$) to reach $z = 0.1$. The DEC method effectively 
    preserves the characteristic doughnut-shaped intensity profile with 
    substantially fewer unknowns.}
    \label{fig:VortexGridMesh}
\end{figure*}

Figure~\ref{fig:VortexGridMesh} presents the output intensity profiles at the normalized propagation distance $z=0.1$ for different grid and mesh resolutions. The top row corresponds to the SIP method on a Cartesian grid, whereas the bottom row shows the DEC solution on the unstructured circular mesh. Evidently, the spectral approach requires fine grids and long computation times to suppress Gibbs errors at the step-index interface, which for this launch is strongly illuminated by near-cutoff odd-azimuthal mode content. As a consequence, preserving the characteristic doughnut-shaped intensity profile of the $m = 1$ vortex becomes computationally demanding for the SIP method. By contrast, the DEC formulation naturally conforms to the material boundary and maintains a high-fidelity representation of the vortex ring using only a small fraction of the degrees of freedom and computational cost.

As a final remark, we emphasize that the propagation distances considered in this work are not an inherent limitation of the DEC-BPM itself, but rather of the current implementation, which employs Neumann boundary conditions. These distances can be extended straightforwardly by incorporating absorbing boundary conditions. To illustrate this point, we performed simulations over substantially longer propagation distances by enlarging the computational domain. For centrally launched fields, this incurs only a modest additional cost. In Supplementary Note 6, we extend the Gaussian beam propagation to five times the original distance using a computational window only 2.5 times as large, while maintaining essentially the same accuracy and relative performance reported in the main text. In contrast, for annular fields launched near the core--cladding interface, such as vortex beams, radiated energy reaches the domain boundary much sooner. Consequently, maintaining the same level of accuracy requires a much larger computational window, making long-distance simulations increasingly uneconomical without the use of absorbing boundary conditions.

\section{Conclusion}
In this work, we introduced a topology-preserving computational framework based on DEC for the efficient simulation of nonlinear wave propagation in complex media. By replacing the continuous domain with a simplicial complex and representing field quantities as discrete differential forms, the governing equations emerge directly from the discrete exterior derivative and Hodge star, without requiring the variational formulation and element-level matrix assembly characteristic of conventional finite-element methods, while at the same time avoiding the geometric restrictions of FFT-based spectral solvers. Unlike Cartesian discretizations, the proposed framework conforms naturally to complex geometries and material interfaces while preserving the topological structure of the underlying differential operators.

Our results demonstrate that this structure-preserving formulation enables accurate and computationally efficient simulations of multiscale nonlinear photonic systems containing sharp refractive-index discontinuities. In these challenging regimes, DEC-BPM resolves complex geometries with substantially fewer degrees of freedom, achieving spectral-level accuracy while delivering computational speedups exceeding two orders of magnitude over conventional spectral methods. Beyond improving computational performance, these gains make routine simulations of previously inaccessible large-scale nonlinear photonic systems computationally practical, reducing runtimes from days to hours and enabling numerical optimization beyond what is currently feasible using simplified models or physical intuition alone.

Another important advantage of the DEC-BPM scheme presented here is its natural extensibility to rigorous absorbing boundary conditions, such as perfectly matched layers (PMLs). In particular, the geometric structure of the discrete operators in the DEC formulation naturally accommodates PMLs through complex stretching of the discrete Hodge stars. Since this stretching modifies the discrete metric rather than any individual operator, the same construction carries over directly to full-vector formulations, in the spirit of exterior complex scaling~\cite{Simon1979}. Incorporating such absorbers is therefore a natural next step toward accurate long-distance propagation in these more general settings.
By contrast, FFT-based algorithms do not enjoy this flexibility. Their efficiency relies on the Fourier basis diagonalizing the constant-coefficient Laplacian, a property that is destroyed by the spatially varying coefficients introduced by a PML transformation. Although absorbing boundary conditions can be incorporated into FFT-based methods through approaches such as complex absorbing potentials~\cite{KosloffKosloff1986,RissMeyer1996,Manolopoulos2002,Muga2004}, these techniques are generally more ad hoc and lack the same geometric and mathematical foundation as PMLs. Consequently, they do not provide comparable guarantees of convergence or a systematic path toward extension to more complex settings, such as waveguides with arbitrary geometries, asymmetric refractive-index profiles, or fully vectorial electromagnetic fields.

More broadly, the present framework establishes discrete exterior calculus as a general computational paradigm for nonlinear wave propagation. In addition to its demonstrated computational efficiency, the structure-preserving formulation naturally accommodates geometry-conforming discretizations, and the systematic construction of higher-order propagation operators from lower-order discrete operators. These capabilities provide a natural pathway toward more general models of nonlinear wave propagation, including ultrashort pulse propagation, vectorial and polarization-dependent formulations, and higher-order nonlinear effects in complex photonic structures. Moreover, because the discrete operators developed here constitute the same algebraic building blocks required for more general electromagnetic models, the framework can be extended systematically while retaining its geometric and topological structure. We anticipate that these capabilities will enable the simulation, optimization, and inverse design of increasingly sophisticated nonlinear photonic devices, including photonic crystal fibres, multicore fibres, multimode waveguides, and integrated nonlinear photonic platforms.


\begin{thebibliography}{26}

\bibitem{weideman1986split}
Weideman, J. A. C. \& Herbst, B. M. Split-step methods for the solution of the nonlinear Schr\"odinger equation. \emph{SIAM J. Numer. Anal.} \textbf{23}, 485 (1986).

\bibitem{TahaAblowitz1984}
Taha, T. R. \& Ablowitz, M. I. Analytical and numerical aspects of certain nonlinear evolution equations. II. Numerical, nonlinear Schr\"odinger equation. \emph{J. Comput. Phys.} \textbf{55}, 203 (1984).

\bibitem{trefethen2000spectral}
Trefethen, L. N. \emph{Spectral Methods in MATLAB} (SIAM, 2000).

\bibitem{Zouraris2001}
Zouraris, G. E. On the convergence of a linear two-step finite element method for the nonlinear Schr\"odinger equation. \emph{ESAIM Math. Model. Numer. Anal.} \textbf{35}, 389 (2001).

\bibitem{li2025efficient}
Li, P. \& Zhang, Z. Efficient finite element methods for semiclassical nonlinear Schr\"odinger equations with random potentials. \emph{ESAIM Math. Model. Numer. Anal.} \textbf{59}, 3249 (2025).

\bibitem{Chen2020Eff}
Chen, J., Li, S. \& Zhang, Z. Efficient multiscale methods for the semiclassical Schr\"odinger equation with time-dependent potentials. \emph{Comput. Methods Appl. Mech. Eng.} \textbf{369}, 113232 (2020).

\bibitem{deschamps1981}
Deschamps, G. A. Electromagnetics and differential forms. \emph{Proc. IEEE} \textbf{69}, 676 (1981).

\bibitem{katz1985differential}
Katz, V. J. Differential forms---Cartan to de Rham. \emph{Arch. Hist. Exact Sci.} \textbf{33}, 321 (1985).

\bibitem{flanders1989}
Flanders, H. \emph{Differential Forms with Applications to the Physical Sciences} 2nd edn (Dover Publications, 1989).

\bibitem{bossavit1998computational}
Bossavit, A. \emph{Computational Electromagnetism: Variational Formulations, Complementarity, Edge Elements} (Academic Press, 1998).

\bibitem{Bossavit1999}
Tarhasaari, T., Kettunen, L. \& Bossavit, A. Some realizations of a discrete Hodge operator: a reinterpretation of finite element techniques. \emph{IEEE Trans. Magn.} \textbf{35}, 1494 (1999).

\bibitem{hirani2003discrete}
Hirani, A. N. \emph{Discrete Exterior Calculus}. PhD thesis, California Institute of Technology (2003).

\bibitem{desbrun2005discrete}
Desbrun, M., Hirani, A. N., Leok, M. \& Marsden, J. E. Discrete exterior calculus. Preprint at https://arxiv.org/abs/math/0508341 (2005).

\bibitem{grady2010}
Grady, L. J. \& Polimeni, J. R. Introduction to discrete calculus. In \emph{Discrete Calculus: Applied Analysis on Graphs for Computational Science} 13--89 (Springer, 2010).

\bibitem{Abdrabou2026hybrid}
Abdrabou, A. \& Gomez, L. J. A hybrid DEC-SIE framework for potential-based electromagnetic analysis of heterogeneous media. \emph{J. Comput. Phys.} \textbf{553}, 114726 (2026).

\bibitem{DelaunayHodge}
Hirani, A. N., Kalyanaraman, K. \& VanderZee, E. B. Delaunay Hodge star. \emph{Comput.-Aided Des.} \textbf{45}, 540 (2013).

\bibitem{Voronoi2007}
Dyer, R., Zhang, H. \& M\"oller, T. Voronoi--Delaunay duality and Delaunay meshes. In \emph{Proc. ACM Symposium on Solid and Physical Modeling} 415--420 (ACM, 2007).

\bibitem{whitney1957}
Whitney, H. \emph{Geometric Integration Theory} (Princeton Univ. Press, 1957).

\bibitem{mohamed2016}
Mohamed, M. S., Hirani, A. N. \& Samtaney, R. Comparison of discrete Hodge star operators for surfaces. \emph{Comput.-Aided Des.} \textbf{78}, 118 (2016).

\bibitem{lohi2021}
Lohi, J. \& Kettunen, L. Whitney forms and their extensions. \emph{J. Comput. Appl. Math.} \textbf{393}, 113520 (2021).

\bibitem{Simon1979}
Simon, B. The definition of molecular resonance curves by the method of exterior complex scaling. \emph{Phys. Lett. A} \textbf{71}, 211 (1979).

\bibitem{KosloffKosloff1986}
Kosloff, R. \& Kosloff, D. Absorbing boundaries for wave propagation problems. \emph{J. Comput. Phys.} \textbf{63}, 363 (1986).

\bibitem{RissMeyer1996}
Riss, U. V. \& Meyer, H.-D. Investigation on the reflection and transmission properties of complex absorbing potentials. \emph{J. Chem. Phys.} \textbf{105}, 1409 (1996).

\bibitem{Manolopoulos2002}
Manolopoulos, D. E. Derivation and reflection properties of a transmission-free absorbing potential. \emph{J. Chem. Phys.} \textbf{117}, 9552 (2002).

\bibitem{Muga2004}
Muga, J. G., Palao, J. P., Navarro, B. \& Egusquiza, I. L. Complex absorbing potentials. \emph{Phys. Rep.} \textbf{395}, 357 (2004).

\bibitem{dormand1980family}
Dormand, J. R. \& Prince, P. J. A family of embedded Runge--Kutta formulae. \emph{J. Comput. Appl. Math.} \textbf{6}, 19 (1980).

\end{thebibliography}
\end{document}